\documentclass[iop,apj,tighten]{emulateapj} 
\usepackage{apjfonts}

\usepackage{amsmath,amssymb,amsfonts}
\usepackage{mathrsfs}
\usepackage{graphicx}
\usepackage{subfigure}
\usepackage{color}

\newcommand{\be}{\begin{equation}}
\newcommand{\ee}{\end{equation}}
\newcommand{\omhat}{\hat{\Omega}} 
\newcommand{\phat}{\hat{p}}
\newcommand{\hplus}{h_+}
\newcommand{\hcross}{h_{\times}}

\newcommand{\lp}{\left(}
\newcommand{\rp}{\right)}
\newcommand{\bb}{\begin{bmatrix}}
\newcommand{\eb}{\end{bmatrix}}

\begin{document}

\title{NANOGrav Limits on Gravitational Waves from Individual \\Supermassive Black Hole Binaries in Circular Orbits}

\author{ 
Z.~Arzoumanian\altaffilmark{1}, 
A.~Brazier\altaffilmark{2}, 
S.~Burke-Spolaor\altaffilmark{3},
S.~J.~Chamberlin\altaffilmark{4},
S.~Chatterjee\altaffilmark{2},
J.~M.~Cordes\altaffilmark{2}, 
P.~B.~Demorest\altaffilmark{5},
X.~Deng\altaffilmark{6},
T.~Dolch\altaffilmark{2},
J.~A.~Ellis\altaffilmark{4},
R.~D.~Ferdman\altaffilmark{7}, 
N.~Garver-Daniels\altaffilmark{8},
F.~Jenet\altaffilmark{9}, 
G.~Jones\altaffilmark{10},
V.~M. Kaspi\altaffilmark{7}, 
M.~Koop\altaffilmark{6},
M.~T.~Lam\altaffilmark{2},
T.~J.~W.~Lazio\altaffilmark{11},
A.~N.~Lommen\altaffilmark{12}, 
D.~R.~Lorimer\altaffilmark{8},
J.~Luo\altaffilmark{9},
R.~S.~Lynch\altaffilmark{7},
D.~R.~Madison\altaffilmark{2},
M.~McLaughlin\altaffilmark{8}, 
S.~T.~McWilliams\altaffilmark{8},
D.~J.~Nice\altaffilmark{13}, 
N.~Palliyaguru\altaffilmark{8},
T.~T.~Pennucci\altaffilmark{14},
S.~M.~Ransom\altaffilmark{5}, 
A.~Sesana\altaffilmark{15},
X.~Siemens\altaffilmark{4}, 
I.~H.~Stairs\altaffilmark{16}, 
D.~R.~Stinebring\altaffilmark{17},
K.~Stovall\altaffilmark{18},
J.~Swiggum\altaffilmark{8},
M.~Vallisneri\altaffilmark{11},
R.~van Haasteren\altaffilmark{11,19},
Y.~Wang\altaffilmark{9},
W.~W.~Zhu\altaffilmark{15}}
\collaboration{NANOGrav Collaboration}


\altaffiltext{1}{Center for Research and Exploration in Space Science and Technology and X-Ray Astrophysics Laboratory, NASA Goddard Space Flight Center, Code 662, Greenbelt, MD 20771, USA}
\altaffiltext{2}{Department of Astronomy, Cornell University, Ithaca, NY 14853, USA}
\altaffiltext{3}{California Institute of Technology, Pasadena, CA 91125, USA}
\altaffiltext{4}{Center for Gravitation, Cosmology and Astrophysics, Department of Physics, University of Wisconsin-Milwaukee, P.O. Box 413, Milwaukee, WI 53201, USA}
\altaffiltext{5}{National Radio Astronomy Observatory, 520 Edgemont Road, Charlottesville, VA 22903, USA}
\altaffiltext{6}{Department of Astronomy and Astrophysics, Pennsylvania State University, University Park, PA 16802, USA}
\altaffiltext{7}{Department of Physics, McGill University, 3600  University St, Montreal, QC H3A 2T8, Canada}
\altaffiltext{8}{Department of Physics, West Virginia University, P.O. Box 6315, Morgantown, WV 26505, USA}
\altaffiltext{9}{Center for Gravitational Wave Astronomy, University of Texas at Brownsville, Brownsville, TX 78520, USA}
\altaffiltext{10}{Department of Physics, Columbia University, New York, NY 10027 USA}
\altaffiltext{11}{Jet Propulsion Laboratory, California Institute of Technology, 4800 Oak Grove Drive, Pasadena, CA 91106, USA}
\altaffiltext{12}{Department of Physics and Astronomy, Franklin and Marshall College, P.O. Box 3003, Lancaster, PA 17604, USA}
\altaffiltext{13}{Department of Physics, Lafayette College, Easton, PA 18042, USA}
\altaffiltext{14}{University of Virginia, Department of Astronomy, PO Box 400325, Charlottesville, VA, 22904}
\altaffiltext{15}{Max-Planck-Institut f{\"u}r Gravitationsphysik, Albert Einstein Institut, Am M{\"u}lenber 1, 14476 Golm, Germany}
\altaffiltext{16}{Department of Physics and Astronomy, University of British Columbia, 6224 Agricultural Road, Vancouver, BC V6T 1Z1, Canada}
\altaffiltext{17}{Department of Physics and Astronomy, Oberlin College, Oberlin, OH 44074, USA}
\altaffiltext{18}{Physics and Astronomy Dept., University of New Mexico, Albuquerque, NM, USA}
\altaffiltext{19}{Einstein Fellow}

\begin{abstract} 
The North American Nanohertz Observatory for Gravitational Waves (NANOGrav) project currently observes 43 pulsars using the Green Bank and Arecibo radio telescopes. In this work we use a subset of 17 pulsars  timed for a span of roughly five years (2005--2010). We analyze these data using standard pulsar timing models, with the addition of time-variable dispersion measure and frequency-variable pulse shape terms. Within the timing data, we perform a search for continuous gravitational waves from individual supermassive black hole binaries in circular orbits using robust frequentist and Bayesian techniques. We find that there is no evidence for the presence of a detectable continuous gravitational wave; however, we can use these data to place the most constraining upper limits to date on the strength of such gravitational waves. Using the full 17 pulsar dataset we place a 95\% upper limit on the sky-averaged strain amplitude of $h_0\lesssim 3.8\times 10^{-14}$ at a frequency of 10 nHz. Furthermore, we place 95\% \emph{all sky} lower limits on the luminosity distance to such gravitational wave sources finding that the $d_L \gtrsim 425$ Mpc for sources at a frequency of 10 nHz and chirp mass $10^{10}{\rm M}_{\odot}$. We find that for gravitational wave sources near our best timed pulsars in the sky, the sensitivity of the pulsar timing array is increased by a factor of $\sim$4 over the sky-averaged sensitivity. Finally we place limits on the coalescence rate of the most massive supermassive black hole binaries.

\end{abstract}

\maketitle

\section{Introduction}

The direct detection of Gravitational Waves (GWs) is a major goal of experimental physics and astrophysics. One of the most promising means of detecting GWs is through the precise timing of an array of millisecond pulsars (MSPs). The concept of a pulsar timing array (PTA) was first conceived of over two decades ago \citep{saz78,det79,hd83,r89,fb90}. Twenty years later three main PTAs are in full operation around the world: the North American Nanohertz Observatory for Gravitational waves \citep[NANOGrav;][]{jfl+09}, the Parkes Pulsar Timing Array \citep[PPTA;][]{m08}, and the European Pulsar Timing Array \citep[EPTA;][]{jsk+08}. The three PTAs collaborate to form the International Pulsar Timing Array \citep[IPTA;][]{haa+10} which will result in increased sensitivity to GWs through more data and longer time-spans than any single PTA.

PTAs are most sensitive to GWs with frequencies in the nanohertz regime (i.e., $10^{-9}$ Hz -- $10^{-7}$ Hz). Potential sources of GWs in this frequency range include supermassive black hole binary systems (SMBHBs) \citep{svc08}, 
cosmic (super)strings \citep{Olmez:2010bi}, inflation \citep{sa79}, and a first order phase transition at the QCD scale \citep{ccd+10}.
The community has thus far focused mostly on stochastic backgrounds produced by these sources; however,  sufficiently nearby 
individual SMBHBs may produce detectable continuous waves with periods on the order of years for masses in the range $10^8 {\rm M}_{\odot}$--$10^{10}{\rm M}_{\odot}$ \citep{wl03,svv09,sv10}. Several upper limits have been placed on the strength of the stochastic background \citep{ktr94, jhs+06,vhj+11, dfg+12, src+13} and continuous waves \citep{jll+04, yhj+10} but no successful detection has yet been made. 

In this paper we will use current-generation frequentist \citep{esc12} and Bayesian \citep{e13} data analysis pipelines to compute upper limits on the strain amplitude of continuous GWs from SMBHBs in circular orbits. We make use of the 5-year, 17 pulsar data set obtained as part of the NANOGrav project \citep{dfg+12}. In Section \ref{sec:obs} we briefly review the radio observations and timing analysis. In Section \ref{sec:gws} we describe the signal model used to describe the continuous GWs in the PTA band. In Section \ref{sec:search} we describe, in detail, the time domain likelihood function, the noise model, and the frequentist and Bayesian search pipelines. In Section 5 we apply our search and upper limit pipelines to the NANOGrav dataset and report our findings. In section \ref{sec:conclusions} we summarize our results. In the Appendices we derive the form of the frequency evolution of SMBHBs, and give full details on the computational implementation of our Bayesian code.

\section{Observations and Timing Analysis}
\label{sec:obs}

The observational data used for this analysis are the same as those
presented by \citet{dfg+12}; the reader is referred to that paper for a
detailed description of the observations and timing analysis.  Here we
present a brief review of the relevant features of the data set.  The
timing data used here were acquired during 2005--2010 using two radio
telescopes, the 305-m Arecibo telescope, and the 100-m Robert C. Byrd
Green Bank Telescope (GBT).  A total of 17 pulsars (8 at Arecibo, 10 at
the GBT, with J1713$+$0747 observed by both telescopes) were monitored
using a typical observational cadence of 4--6 weeks between sessions.
At each observing epoch, every pulsar was observed using two separate
receiver systems operating at widely separated radio frequencies ranging
from 327~MHz to 2.3~GHz.  The typical observation length was 30 minutes
per pulsar per receiver.  All data were recorded using the identical ASP
(at Arecibo) and GASP (at the GBT) pulsar backend systems
\citep{d07}.  These systems processed a typical radio bandwidth
of 64~MHz using real-time coherent dedispersion and pulse period
folding, resulting in 2048-bin full-Stokes pulse profiles averaged over
1--3~minutes in 4~MHz channels.

Pulse profile calibration, integration, and time of arrival (TOA)
determination was done using standard techniques via the
PSRCHIVE\footnote{\texttt{http://psrchive.sourceforge.net}} software
package \citep{hvsm04}.  For each pulsar all profiles in a given epoch
were integrated in time to form a single set of profiles across
radio frequency.  From these, TOAs were measured {\it separately} in
each 4~MHz radio frequency channel.  This resulted in a set of
$\sim$20--30 multi-frequency TOAs at each epoch, or $\sim$500--2000 TOAs
total for each pulsar over the full data set.  Before searching for the
presence of GW in these data, the rotational, orbital, astrometric and
interstellar medium properties specific to each pulsar -- effects
collectively known as the timing model -- must first be determined from
the TOA data.  For this we analyzed the TOAs using both the
TEMPO\footnote{\texttt{http://tempo.sourceforge.net}} and
TEMPO2\footnote{\texttt{http://tempo2.sourceforge.net}} \citep{hem06}
timing software packages and obtained identical results with both.
Notable features of the timing models used here include: Spin frequency
and spin-down rate, but no higher frequency derivatives, were fit for
all pulsars;  all five astrometric parameters (sky position, proper
motion and parallax) were fit for all pulsars\footnote{Parallax was not fit for in PSR J1640$+$2224.}; time-variable dispersion
measure (DM); was included by fitting for an {\it independent} DM value at
each epoch, using the multi-frequency TOAs;\footnote{Models for pulsars
J1853$+$1308, J1910$+$1256 and B1953$+$29 did not include DM variation
measurement as only single-frequency data were available for these.}
intrinsic profile shape evolution with frequency; was included as a
constant-in-time offset for each frequency channel, and Keplarian and relativistic orbital elements, as appropriate for pulsars in binary systems.  
All TOA data and final timing solutions for this data set are publicly available
online.\footnote{\texttt{http://data.nanograv.org}}

\section{GWs From Supermassive Black Hole Binaries}
\label{sec:gws}

Pulsar timing residuals are defined as the difference between observed times-of-arrival (TOAs) of radio pulses and a deterministic timing model. In this section we review the form of the residuals induced by SMBHBs consisting of non-spinning black holes in a circular orbit \citep[e.g.,][]{w87} and introduce our notation. Spin effects are not likely to play any measurable role in the orbital dynamics \citep{sv10} and eccentric systems \citep{rs11, s13} will be addressed in a future work. The GW 
is a metric perturbation to flat space-time defined in terms of its two polarizations as
\be
h_{ab}(t,\omhat)=e_{ab}^+(\omhat)\hplus(t,\omhat)+e_{ab}^{\times}(\omhat)\hcross(t,\omhat),
\ee
where $\omhat$ is the unit vector pointing from the GW source to the Solar
System Barycenter (SSB), $\hplus$, $\hcross$ and $e_{ab}^A$ ($A=+, \times$) are the  
polarization amplitudes and polarization tensors, respectively. The 
polarization tensors can be converted to the  
SSB by the following transformation. Following \citet{w87} we write
\begin{align}
e_{ab}^+(\omhat)&=\hat{m}_a\hat{m}_b-\hat{n}_a\hat{n}_b,\\
e_{ab}^{\times} (\omhat)&=\hat{m}_a\hat{n}_b+\hat{n}_a\hat{m}_b,
\end{align}
where
\begin{align}
\omhat &=-(\sin\theta\cos\varphi)\hat{x}-(\sin\theta\sin\varphi)\hat{y}-(\cos\theta)\hat{z},\\
\hat{m} &=-(\sin\varphi)\hat{x}+(\cos\varphi)\hat{y},\\
\hat{n} &=-(\cos\theta\cos\varphi)\hat{x}-(\cos\theta\sin\varphi)\hat{y}+(\sin\theta)\hat{z},
\end{align}
where $\hat{x}$, $\hat{y}$, and $\hat{z}$ are the usual Cartesian coordinate unit vectors. In this coordinate system, $\theta=\pi/2-\delta$ and $\varphi=\alpha$ are the polar and azimuthal angles of the source, respectively, where $\delta$ 
and $\alpha$ are declination and right ascension in the usual equatorial coordinates, where the north celestial pole is in the $\hat{z}$ direction and the vernal equinox is in the $\hat{x}$ direction.

We will write our GW induced pulsar timing residuals in the following form:
\be
s(t,\omhat)=F^{+}(\omhat)\Delta s_{+}(t)+F^{\times}(\omhat)\Delta s_{\times}(t),
\ee
where 
\be
\Delta s_{A}(t)=s_{A}(t_{p})-s_{A}(t_{e}),
\ee
and $t_{e}$ and $t_{p}$ are the times at which the GW passes the Earth\footnote{Technically, this is the time that the GW passes the SSB, however, following convention we will label this as the \emph{Earth} time and will later refer to the \emph{Earth}-term, keeping in mind that, in practice, all variables are referenced to the SSB.} and pulsar, respectively, and
the index $A\in\{+,\times\}$ labels polarizations. 
The functions $F^{A}(\omhat)$ are known as antenna pattern functions and are defined by
\begin{align}
F^{+}(\omhat)&=\frac{1}{2}\frac{(\hat{m}\cdot\phat)^{2}-(\hat{n}\cdot\phat)^{2}}{1+\omhat\cdot\phat}\\
F^{\times}(\omhat)&=\frac{(\hat{m}\cdot\phat)(\hat{n}\cdot\phat)}{1+\omhat\cdot\phat},
\end{align}
where $\phat$ is the unit vector pointing from the Earth to the pulsar. Also, from geometry we can write\footnote{Here we use geometrized units where $G=c=1$.}
\be
\label{eq:pTime}
t_{p}=t_{e}-L(1+\omhat\cdot\phat)
\ee
where $L$ is the distance to the pulsar. Given these definitions, we can write the GW contributions to the timing residuals at 0-PN (Post-Newtonian) order as \citep{w87,cc10}
\begin{align}
\begin{split}
\label{eq:rplus}
s_{+}(t)&=\frac{\mathcal{M}^{5/3}}{d_L\omega(t)^{1/3}}\Big[-\sin[2\Phi(t)](1+\cos^{2}\iota)\cos2\psi\\
&-2\cos[2\Phi(t)]\cos\iota\sin2\psi\Big]
\end{split}\\
\begin{split}
\label{eq:rcross}
s_{\times}(t)&=\frac{\mathcal{M}^{5/3}}{d_L\omega(t)^{1/3}}\Big[-\sin[2\Phi(t)](1+\cos^{2}\iota)\sin2\psi\\
&+2\cos[2\Phi(t)]\cos\iota\cos2\psi\Big],
\end{split}
\end{align}
where $\psi$ is the GW polarization angle and $\iota$ is the inclination angle of the SMBHB. The orbital phase and 
frequency of the SMBHB are 
\be
\label{eq:phit}
\Phi(t)=\Phi_{0}+\frac{1}{32\mathcal{M}^{5/3}}\lp\omega_{0}^{-5/3}-\omega(t)^{-5/3}\rp
\ee
and
\be
\label{eq:worb}
\omega(t)=\omega_0\lp 1-\frac{256}{5}\mathcal{M}^{5/3}\omega_0^{8/3}t \rp^{-3/8}.
\ee
where $\Phi_{0}$ and $\omega_{0}$ are the initial values at the time of our first observation, the chirp mass is defined by $\mathcal{M}=(m_{1}m_{2})^{3/5}/(m_{1}+m_{2})^{1/5}$, where $m_{1}$ and $m_{2}$ are the masses of the two SMBHs, and $d_L$ is the luminosity distance to the SMBHB source. See Appendix \ref{sec:freq_ev} for a more complete derivation of the frequency evolution of the binary including important approximations that can be made. For our purposes here we will just note that orbital frequency evolution over our observing time span is very unlikely while frequency evolution over the earth-pulsar light travel time is almost certain for sources with reasonably large chirp masses (i.e., $\mathcal{M}\sim 3\times 10^{8}{\rm M}_{\odot}$) \citep[see e.g.,][]{sv10}. We can relate the GW frequency to the orbital frequency of the binary by $\omega_{\rm gw}=2\omega_0$ for circular orbits. Note that we use the observed redshifted values. For example, the chirp mass and orbital angular frequency in the rest frame are $\mathcal{M}_{r}=\mathcal{M}/(1+z)$ and $\omega_{r}=\omega_{0}(1+z)$, respectively, where $z$ is the cosmological redshift. 

From the signal model presented above, we see that our parameter space is 8 dimensional and the continuous wave parameter space vector is
\be
\vec\lambda_0 = \{ \theta, \varphi, \Phi_0, \psi, \iota, \mathcal{M}, d_L, \omega_0\}.
\ee
However, because typical pulsar distance uncertainties are on the order of tens of percent \citep{vwc+12}, in order to attain phase coherence in our search algorithm, we must allow the pulsar distance to vary as a search parameter as well. Henceforth, we will adopt the notation that $\vec\lambda_{\alpha} = \{  \vec\lambda_0, L_{\alpha}\}$, where $L_{\alpha}$ is the distance to the $\alpha$th pulsar, in order to denote the fact that the pulsar distance is a search parameter. The above parameter set represents the default parameters used in our search; however, when setting upper limits we wish to parameterize the upper limit in terms of the inclination averaged strain amplitude 
\be
h_0 = 4 \sqrt{\frac{2}{5}}\frac{\mathcal{M}^{5/3}(\pi f_{\rm gw})^{2/3}}{d_L}.
\label{eq:strain_amp}
\ee
Since the luminosity distance, $d_L$ is only a scale parameter we use $h_0$ as a free parameter in the waveform instead of luminosity distance when computing upper limits.

\section{Search Techniques}
\label{sec:search}

\subsection{Likelihood Function for PTAs}

Following \cite{vhl12} and \cite{esvh13} we write the measured pulsar timing residuals in the linear approximation as
\be
\delta t=M\delta\boldsymbol{\xi}+n+s,
\ee
where $\delta t$ is a vector of length $N_{\rm TOA}$ representing the timing residuals for a single pulsar, $M$ is the design matrix (an $N_{\rm TOA}\times m$ matrix composed of the functional form of the linearized timing model (column) evaluated at each TOA (row)), $\delta\boldsymbol{\xi}$ is a vector of length $m$ representing the parameter offset between the true pulsar timing parameters and our best fit parameters, $n$ and $s$ are length $N_{\rm TOA}$ vectors representing the noise present in the TOAs (radiometer noise, red noise, etc.) and our continuous GW signal, respectively.  In practice, the noise in pulsar timing residuals is non-Gaussian due to interstellar medium scintillation effects which are manifest through a time varying pulse intensity. Nonetheless, the noise in each residual is modeled very well by a Gaussian with zero mean and standard deviation equal to the uncertainty on the TOA. In other words, the noise in the \emph{weighted} (by the individual TOA errors) residuals is very well approximated as a Gaussian. As will be detailed in the next section, we include these error bar weights in our noise covariance matrix. Thus, assuming $n$ is Gaussian\footnote{Note that here Gaussian noise simply means that the data obey a multivariate Gaussian probability distribution function. This does not mean that we assume the data is white.}, we can write the likelihood function for a single pulsar as
\be
p(\delta t|\delta\boldsymbol{\xi},\vec\phi,\vec\lambda)=\frac{\exp\left[- \frac{1}{2} \lp \delta t -s -M\delta\boldsymbol{\xi} \rp^T C^{-1}\lp \delta t -s -M\delta\boldsymbol{\xi} \rp\right]}{\sqrt{(2\pi)^{N_{\rm TOA}}\det C}},
\label{eq:like}
\ee
where $\vec\phi$ are parameters that describe the noise in the pulsar residuals, $\vec\lambda$ the parameters that characterize the continuous GW signal\footnote{We have dropped the $\alpha$ subscript on the parameter vector here as we are only considering a single pulsar.} and $C$ is the covariance matrix of the noise. It was shown in \cite{vhl12} that this likelihood function can be marginalized over the timing model parameters $\delta\boldsymbol{\xi}$ to obtain
\be
p(\delta t|\vec\phi,\vec\lambda)=\frac{\exp\left[- \frac{1}{2} \lp \delta t -s \rp^T G(G^T CG)^{-1} G^T\lp \delta t -s\rp\right]}{\sqrt{(2\pi)^{(N_{\rm TOA}-m)}\det(G^TCG)}},
\ee
where $G$ is an $N_{\rm TOA}\times (N_{\rm TOA}-m)$ matrix with $N_{\rm TOA}$ the number of TOAs and $m$ the number of fitted parameters in the timing model. The derivation of the $G$-matrix approach can be found in \cite{vhl12} and will not be explored here. The matrix $G^T$ is a projection operator that projects our data $\delta t$ onto the null space of $M$, that is, it projects the data into a subspace orthogonal to the linearized timing model. In the timing analysis used here, DM variation and profile frequency evolution effects are part of the timing model, and these terms are included when constructing the $G$ matrix. In this way we have fully taken into account the timing model fitting procedure.

For this work we will assume that  the residuals between pulsars are uncorrelated. In other words, we are assuming that the stochastic GW background will be negligible compared to the intrinsic noise in each pulsar. In general this is not likely to be a good assumption when we would expect a detection of a single GW source.  Furthermore, terrestrial clock errors \citep{hcm+12} and errors in solar system ephemerides \citep{chm+10}\footnote{Note that current uncertainties in the ephemerides are small enough that they will likely not pose any problems for GW analyses.} can also cause correlations between residuals from different pulsars,  with however different angular correlation properties on the sky than are expected from GWs. The effects of omitting the correlations in the likelihood function are unknown and will be the subject of future work. Under these assumptions, the likelihood function for the full PTA can be written as
\be
p(\delta t|\vec\lambda)=\prod_{\alpha=1}^{N_{\rm psr}}p(\delta t_{\alpha}|\vec\lambda_{\alpha}),
\ee
where $\delta t_{\alpha}$ and $\vec\lambda_{\alpha}$ and the residuals and model parameters for the $\alpha$th pulsar, respectively and $\vec\lambda$ is the full CW parameter vector including pulsar distances for all pulsars.
In cases where we fix the noise values, we can write the log-likelihood ratio of a model with a single continuous GW to a model with just noise as
\be
\label{eq:lnlike}
\ln\,\Lambda=\sum_{\alpha}^{N_{\rm psr}}\left[ \lp \delta t_{\alpha}|s(\vec\lambda_{\alpha})\rp-\frac{1}{2}\lp s(\vec\lambda_{\alpha})|s(\vec\lambda_{\alpha})\rp\right],
\ee
where the inner product between two time-series $x$ and $y$ is
\be
(x|y)=x^TG(G^T CG)^{-1} G^Ty.
\ee
In the remainder of the paper we will refer to the signal-to-noise ratio in the following form
\be
\rho=\sqrt{2\langle \ln\,\Lambda\rangle}= \lp \sum_{\alpha}^{N_{\rm psr}} \lp s(\vec\lambda_{\alpha})| s(\vec\lambda_{\alpha}) \rp \rp^{1/2},
\label{eq:snr}
\ee
where the brackets denote the expectation value over many noise realizations.

\subsection{Noise Model}
\label{sec:noisemodel}

In the above section we have derived the likelihood function used for our analysis; however, we have not specified the form of the noise covariance matrix $C$. Previous Bayesian analysis schemes \citep{hlm+09, vl10, vhj+11, esc12, vhl12, vh13, esvh13, e13} have used a power-law red noise model and an EFAC (constant multiplier on the TOA uncertainties) and EQUAD (additional Gaussian white noise added in quadrature to EFAC noise) parameters to describe the white noise, with a covariance matrix of the form
\be
C = E^2W + \mathcal{Q}^2\mathbb{I} + C^{\rm red}(A_{\rm red}, \gamma_{\rm red}),
\ee
where $E$ is the EFAC parameter, $W={\rm diag}\{\sigma_i^2\}$, with $\sigma_i$ the errorbar on the $i$th TOA, $\mathcal{Q}$ is the EQUAD parameter and $C_{\rm red}$ is an analytic expression of the red noise amplitude $A_{\rm red}$ and spectral index $\gamma_{\rm red}$. It is worth noting that we use no EFAC or EQUAD parameters in our pulsar timing model fit but instead include them directly in our noise model. The EFAC is simply a parameter that quantifies any additional uncertainty in the TOA uncertainties and the EQUAD parameter quantifies any additional white noise that is not related to the formal TOA uncertainties. In principle, a different EFAC value should be used for each pulsar timing backend as this parameter is related to intrinsic receiver noise; however, in this 5-year NANOGrav dataset, only one backend per telescope was used\footnote{PSR J1713+0747 is observed at both Arecibo and GBT; however, we find that there is very little difference in the measured EFAC parameters for the two telescopes.}. Therefore, we are justified in only using one EFAC parameter per pulsar. This noise model is quite general and works well for many pulsars; however, the size of the matrices is quite large (on the order of $10^3\times 10^3$) and inversion is a large bottleneck in the analysis pipelines. Furthermore, current NANOGrav observing schemes produce large sets of multifrequency observations that are essentially simultaneous. One may be tempted to simply perform a weighted average of the TOAs and work with the new reduced datasets but in the Bayesian scheme we must marginalize over the timing model parameters analytically and it is unclear how to carry out this process for epoch-averaged TOAs. Because of this, we have developed a framework to essentially work backward from the marginal likelihood to derive a nearly exact averaging scheme. First we re-write our noise covariance matrix
\be
C = N + U\tilde{\mathcal{C}}U^T,
\ee
where $\tilde{\mathcal{C}}$ is a $q\times q$ reduced covariance matrix with $q$ the number of epochs\footnote{Here we have defined an epoch to be one day.} in our dataset, $N$ is a white noise covariance matrix of the EFAC and EQUAD terms, and $U$ is the ``exploder" matrix that maps epochs (columns) to the full set of TOAs (rows).
If we now make use of this new formalism, the likelihood function is then
\be
p(\delta t|\vec\phi,\vec\lambda) = \frac{\exp\left[ -\frac{1}{2} \lp(\delta t-s)^T \tilde{N}^{-1} (\delta t-s) - d^T \Sigma^{-1}d\rp\right]}{\sqrt{(2\pi)^{n-m}\det(\tilde{\mathcal{C}}) \det (G^TNG) \det (\Sigma)}},
\ee
where we have used the Woodbury Lemma\footnote{$(A+DBE^T)^{-1} = A^{-1}-A^{-1}D(B^{-1}+E^TA^{-1}D)^{-1}E^TA^{-1}$ and $|A+DBE^T|=|A||B||B^{-1} + E^TA^{-1}D|$} to compute the inverse and determinant of $C$,  $\tilde{N}^{-1} = G\lp  G^T N G \rp^{-1} G^T$, $d = U^T\tilde{N}^{-1}(\delta t-s)$, and $\Sigma = \lp \tilde{\mathcal{C}}^{-1} + U^T\tilde{N}^{-1}U \rp$. Note, that $d$ here are essentially daily averaged residuals. For NANOGrav datasets the number of epochs per pulsar is on the order of 30--100, while the total number of TOAs per pulsar is on the order of $10^3$, thus the inversions (here $N$ is diagonal and $\tilde{N}^{-1}$ can be pre-computed, thus the only dense matrix inversion is $\Sigma^{-1}$) required in this likelihood function scale as $q^3$ as opposed to $n^3$, resulting in computational speedups of several orders of magnitude. Furthermore, the epoch-averaged covariance matrix $\tilde{\mathcal{C}}$ can take on several forms depending on the red noise model used; however, as long as it is a slowly varying function of the TOAs (i.e., a truly red process) then this formalism is completely valid.

In order to attain further computational speedups and to gain more control over the low frequency component of our noise model we make use of the methods described in \cite{lha+13}, but now applied to a single pulsar instead of the full PTA. This method relies on explicitly splitting up the red and white components of the residuals, so that the residuals are now written as
\be
\delta t = M\delta\boldsymbol\xi + n_{\rm white}+ n_{\rm red} + s,
\ee
where $n_{\rm white}$ and $n_{\rm red}$ are the white and red components of the residuals, respectively. It is possible to expand the red noise piece in a Fourier series
\be
n_{\rm red} = \sum_{j=1}^{N_{\rm mode}}\left [ a_j \sin\lp \frac{2\pi j t}{T}\rp + b_j \cos\lp \frac{2\pi j t}{T}\rp \right] = F\mathbf{a},
\ee
where $\mathbf{a}$ is a vector of the concatenated sine and cosine amplitudes, $T$ is the total time span of the data, and $F$ is a $N_{\rm TOA} \times 2N_{\rm mode}$ matrix with alternating sine and cosine terms with $N_{\rm mode}$ the number of frequencies used. Now, we assume that the underlying ensemble average red noise process is wide-sense stationary and can be completely described by a power-spectrum. Then, by orthogonality, the Fourier coefficients $\mathbf{a}$ will be diagonal with components
\be
\varphi_{ij} = \langle \mathbf{a} \mathbf{a}^T\rangle_{ij} = {\rm diag}(\{\varphi_i\}),
\ee
where the elements of $\varphi$, denoted $\{ \varphi_i \}$ are the coefficients of the theoretical power spectrum of the red noise process in the residuals. If the red noise process is wide-sense stationary, then this relation is always true irrespective of the sampling as all information about the uneven sampling here comes from the Fourier design matrix $F$. Thus, we can write the covariance and epoch-averaged covariance matrices, respectively, as
\begin{align}
C &= N + F\varphi F^T\\
\tilde{\mathcal{C}} & = \tilde{F}\varphi \tilde{F}^T,
\end{align}
where $\tilde{F}$ is a $q\times N_{\rm mode}$ matrix and is constructed in the same manner as $F$ but the epoch-averaged TOAs are used as opposed to the full set of TOAs. As is done in \cite{lha+13}, it is possible to treat each diagonal element of $\varphi$ as a free parameter; however, for this work we choose to parameterize it by a power-law
\be
\varphi_i = \frac{1}{T}\frac{A_{\rm red}^2}{12 \pi^2}\lp \frac{f_i}{f_{\rm yr}} \rp^{3-\gamma_{\rm red}} f_i^{-3},
\ee
where $f_i$ is the $i$th Fourier frequency assuming Nyquist sampling. In general, any Fourier based method with finite length datasets and especially with irregular sampling will suffer from spectral leakage whereby power from the lowest frequencies will leak into the higher frequencies. In effect, this makes Fourier based methods sensitive to the low-frequency cutoff. However, it was shown in \cite{vhl12} that by including the effects of the timing model (specifically the quadratic spin-down in this case) in our analysis acts as a window function that fully removes any sensitivity to the low-frequency cutoff, thereby also removing any spectral leakage. We have done extensive simulations to test this notion and have found no evidence for spectral leakage and no bias in red noise parameter estimation and waveform reconstruction.

In the course of our single pulsar noise analysis \citep{esd+13} we found that the addition of an extra white noise parameter was needed to accurately describe the data. This new white noise term incorporates a correlation among frequency channels (within a given epoch) while still remaining independent of other epochs. In other words, this white noise term accounts for epoch-to-epoch fluctuations as opposed to fluctuations within an epoch. We defer to another paper the inclusion of pulse-jitter noise from pulsar magnetospheric activity but point out that our inferred extra term may be the same as jitter noise known to be present in all well-studied pulsars \citep{cs10}. This parameter is quite easy to incorporate as it is simply an EQUAD like parameter in the epoch-averaged sense, that is
\be
J = U\tilde{J}U^T = \mathscr{J}^2U\mathbb{I}_{q}U^T,
\ee
where $\mathscr{J}$ is our frequency correlated EQUAD parameter and $\mathbb{I}_q$ is the identity matrix in the epoch-averaged space. With this, we have our final noise model with a total covariance matrix of
\be
C = N + U\lp \tilde{F}\varphi \tilde{F}^T + \mathscr{J}^2\mathbb{I}_q \rp U^T,
\ee
and noise parameter vector
\be
\vec\phi = \{E, \mathcal{Q}, \mathscr{J}, A_{\rm red}, \gamma_{\rm red}\}.
\ee
Throughout the remainder of the paper, this noise model is always used for all pulsars.

\subsection{$\mathcal{F}_p$-Statistic}
\label{sec:fpstat}

The $\mathcal{F}_p$-statistic was first derived from the likelihood function of Eq. \eqref{eq:like} in \cite{esc12} (hereafter ESC12) as a ``total-power'' frequentist detection statistic. We will not derive the full expression here (See Appendix \ref{sec:alternatefstat} for an alternate derivation to ESC12), but rather we will explain its functional form and discuss its statistics. First we define the following harmonic basis functions:
\begin{align}
B_{\alpha}^{1}(t)&=\frac{1}{\omega_{0}^{1/3}}\sin(2\omega_{0}t)\\
B_{\alpha}^{2}(t)&=\frac{1}{\omega_{0}^{1/3}}\cos(2\omega_{0}t),
\end{align}
where, again, $\omega_{0}$ is the orbital angular frequency of the SMBHB.  Following ESC12, the $\mathcal{F}_p$-statistic is written as
\be
2\mathcal{F}_p = \sum_{\alpha=1}^{M}P^{i}_{\alpha}Q_{ij}^{\alpha}P_{\alpha}^{j},
\ee
where we have assumed Einstein Summation notation over latin indices, $P^{i}_{\alpha} = (\delta t|B^i_\alpha(t))$, $Q_{ij}^{\alpha}=(B^{\alpha}_{i}|B^{\alpha}_j)$ and the formal sum is over all pulsars in the array. An intuitive way to think of this statistic is a weighted (by the noise power spectral density) sum of the power spectrum of the residual data done in the time domain by making use of a harmonic time domain basis. It was shown in ESC12 that $2\mathcal{F}_p$ follows a chi-squared distribution with $2N_{\rm psr}$ degrees of freedom and non-centrality parameter $\rho^2$ such that
\be
\langle 2\mathcal{F}_p \rangle = 2N_{\rm psr}+ \rho^2.
\ee
\begin{figure}
  \begin{center}
  \includegraphics[scale=0.53]{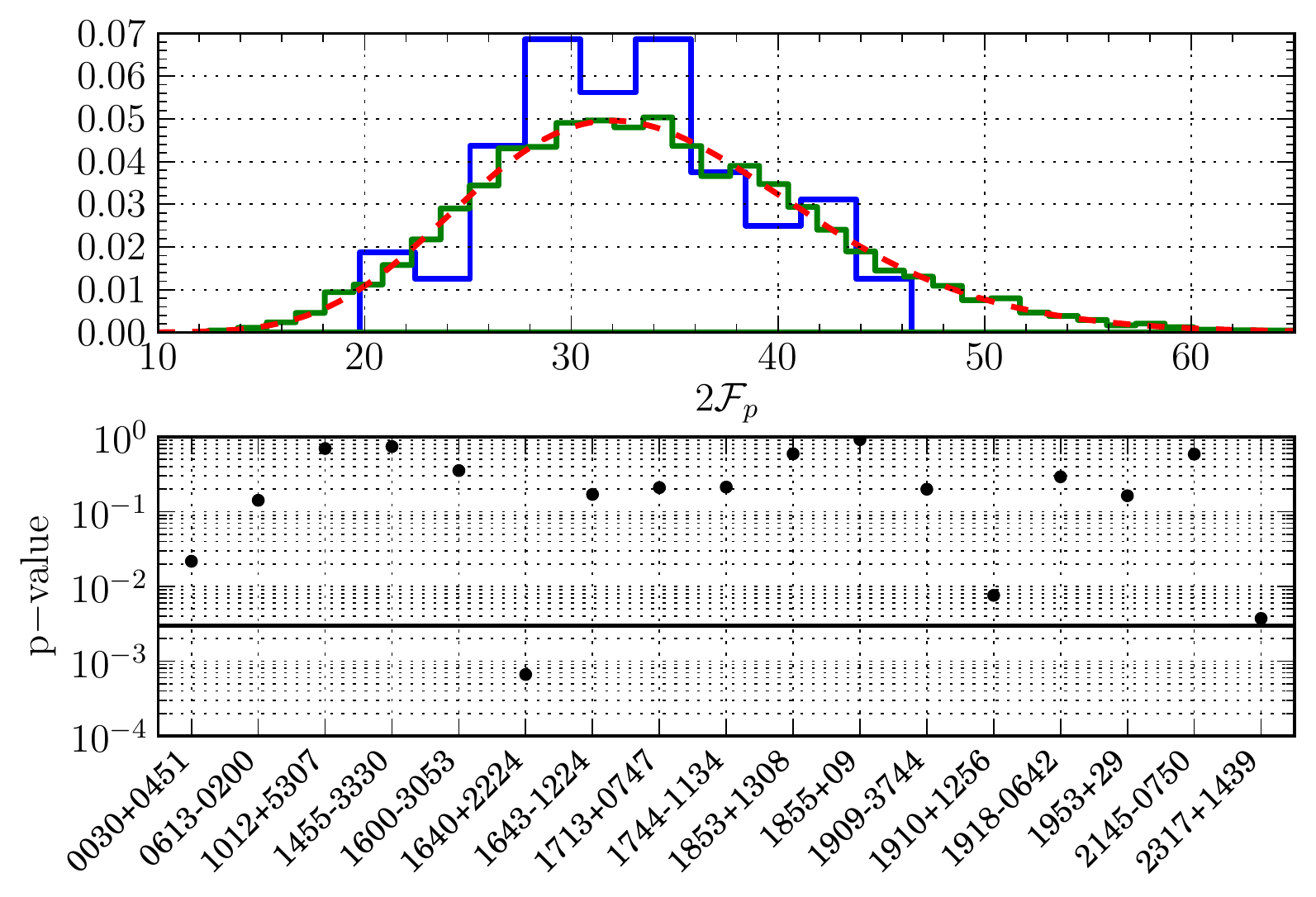}
   \end{center}
  \caption{Histogram of $\mathcal{F}_p$-statistic values (top panel) across all independent frequencies (black(blue) histogram) and for 100,000 realizations of simulated data with noise parameters measured from the real data (gray(green) histogram). The red dashed curve is the probability distribution function for a chi-squared distribution with 34 (i.e., 2$N_{\rm psr}$) degrees of freedom. The lower panel shows the $p$-value from a KS test comparing the $\mathcal{F}_p$-statistic for each pulsar to a chi-squared distribution with 2 degrees of freedom. The solid line represents the 3-sigma threshold for the $p$-value. (Color figure available in online version.)}
  \label{fig:fpstat_pval}
\end{figure}
Figure \ref{fig:fpstat_pval} shows the distribution of the $\mathcal{F}_p$-statistic (top panel) for both real and simulated data as well as as a $p$-value test (bottom panel) for each pulsar, where we compare the single-pulsar $\mathcal{F}_p$ distribution to the expected chi-squared distribution. To compute the $\mathcal{F}_p$-statistic, we have used the maximum a-posteriori noise values obtained in a previous single-pulsar noise analysis to construct the noise covariance matrix. Since we do not have independent realizations of our data, we compute the $\mathcal{F}_p$-statistic for each independent\footnote{Note that the frequencies are not completely independent since our data are irregularly sampled. The frequency bins were chosen here assuming a cadence of two observing sessions per month.} frequency bin and then construct a histogram of the results. If our noise model is a good description of the true noise in our data and there is no GW present in the data then this distribution should follow the correct chi-squared distribution. We see from Figure \ref{fig:fpstat_pval} that the $\mathcal{F}_p$-statistic values do indeed follow a chi-squared distribution with $2N_{\rm psr}$ degrees of freedom. The black(blue) curve in the top panel of Figure \ref{fig:fpstat_pval} shows the aforementioned histogram along with the chi-squared distribution in the dashed gray(red) line. The $p$-value that results from a Kolmogorov-Smirnov (KS) test comparing the $2\mathcal{F}_p$ and chi-squared (with 34 degrees of freedom) distributions is 0.33 showing good agreement between our data and the expected chi-squared distribution. As a cross-check, we have also simulated 100,000 datasets with the measured noise parameters and have evaluated the $\mathcal{F}_p$-statistic for each. This distribution is plotted as a gray(green) histogram in the figure and it is obvious that this distribution follows a chi-squared distribution with 34 degrees of freedom nearly perfectly. We have also performed a similar test but for each pulsar separately. In the lower panel of Figure \ref{fig:fpstat_pval} we carry out the same KS-test mentioned above but now compute the $\mathcal{F}_p$-statistic values for each pulsar individually and then compare to a chi-squared with 2 degrees of freedom. The solid line corresponds to the p-value at which we should reject the null hypothesis that the two distributions are the same with 99.7\% confidence. We see that with the exception of one pulsar, J1640$+$2224, all others lie above this threshold value. This indicates that our noise model for all pulsars except J1640$+$2224 provide a good description of the true noise in the dataset. Better noise models for this pulsar are currently being explored \citep{esd+13} but since our full 17-pulsar $\mathcal{F}$-statistic distribution is totally consistent with the expected chi-squared distribution we just use the standard noise model described in Section \ref{sec:noisemodel}. 

For the detection problem, we are interested in the false-alarm-probability (FAP), that is, the probability that a measured value $\mathcal{F}_p$ exceeds a given threshold $\mathcal{F}_{p,0}$ when no signal is present. From ESC12, the probability distribution of $\mathcal{F}_p$ when the signal is absent is
\be
p_{0}(\mathcal{F}_p)=\frac{\mathcal{F}_p^{n/2-1}}{(n/2-1)!}\exp(-\mathcal{F}_p),
\ee
where $n$ is the number of degrees of freedom ($2N_{\rm psr}$ in this case). The corresponding FAP is then written as
\be
P_{F}(\mathcal{F}_{p,0})=\int_{\mathcal{F}_{p,0}}^{\infty}p_{0}(\mathcal{F}_p)d\mathcal{F}_p
=\exp(-\mathcal{F}_{p,0})\sum_{k=0}^{n/2-1}\frac{\mathcal{F}_{p,0}^{k}}{k!}.
\label{eq:FAP}
\ee
In a search over GW frequencies (the only free parameter in the $\mathcal{F}_p$-statistic) we will incur a trials factor such that the resulting FAP for the search is

\be
P_{F}^{T}(\mathcal{F}_{p,0})=1-\left[ 1-P_{F}(\mathcal{F}_{p,0}) \right]^{N_{f}},
\label{eq:FAPT}
\ee
where $N_f$ is the number of independent frequencies. For this work we place our detection threshold on $\mathcal{F}_p$ such that the corresponding FAP is less than $10^{-4}$. The results of performing this search on the 5-year NANOGrav dataset will be presented in the next section.

\subsection{Bayesian Method}

The Bayesian search pipeline in this work is very similar to that of \cite{e13} (hereafter E13). Here we use an MPI enabled Parallel-Tempered Markov Chain Monte-Carlo (PTMCMC) sampler\footnote{\texttt{https://github.com/jellis18/PAL}} (See E13 and Appendix \ref{sec:jumps} for details on the implementation). In this work we use two ``modes'' of operation for the Bayesian search. The first is the most general in which we evaluate the full likelihood function of Eq. \eqref{eq:like} and allow both the GW parameters, $\vec\lambda$, and the noise model parameters, $\vec\phi$ to vary simultaneously. In principle, this is the more desirable setup as it allows the uncertainty in our noise model to propagate into the measured GW parameters and also accounts for any correlations between the noise and GW parameters. This mode does require significantly more computational power as the number of search parameters in the MCMC is quite large. The total parameter space consists of 8 GW parameters, $N_{\rm psr}$ pulsar distances,  and $5\times N_{\rm psr}$ noise parameters; this comes to 110 parameters for the full 17-pulsar array. 

The second mode is when we fix the noise parameters to their maximum a-posteriori values obtained from a previous single pulsar analysis. All previous GW searches for single sources have been performed in this manner \citep{jll+04,yhj+10, bs12, esc12,pbs+12,e13} which is justified \emph{if} the noise model only contains white noise and the GW signal present in any single dataset is weak. If the noise model contains only white noise, there will be little to no correlation between the GW parameters and the noise parameters, and if the signal is weak then it will not affect the single pulsar noise analysis. However, there is some evidence of red noise in our pulsars and because of the highly varying noise levels among pulsars, it is likely that a \emph{detectable} source would be seen in the best timed pulsars individually. Therefore, this type of Bayesian analysis is not robust and could possibly lead to biased results; nonetheless, we will carry out this mode for comparison purposes in this study. Note that we will have the same problem with the $\mathcal{F}_p$-statistic. Possible methods to ameliorate this problem in fixed-noise searches are being explored and will be the subject of a future paper.

In a Bayesian sense, the detection problem is an application of model selection and the upper limit problem is an application of parameter estimation. The upper limit problem will be discussed in the next section. The model selection technique is very similar to that discussed in E13 and we will only review it here (See Appendix \ref{sec:PTev} for more details). In the Bayesian framework, the data $d$ (here we use $d$ as opposed to the actual residual data $\delta t$ to remain general) are assumed to be fixed and the parameters $\Theta$ that parameterize a hypothesis (or model) $\mathcal{H}$ are assumed to follow a given prior distribution.  In this case, the data are used to update our prior knowledge of the hypothesis $p(\Theta|\mathcal{H})$ via Bayes theorem.
\be
p(\Theta|d,\mathcal{H})=\frac{p(d|\Theta,\mathcal{H})p(\Theta|\mathcal{H})}{p(d|\mathcal{H})},
\ee
where $p(\Theta|d,\mathcal{H})$ is the posterior probability distribution, that is, the probability that the set of parameters $\Theta$ for hypothesis $\mathcal{H}$ could generate the given data $d$. In the above expression $p(d|\Theta,\mathcal{H})$ is the likelihood function, the probability that this dataset $d$ is drawn from a random distribution described by hypothesis $\mathcal{H}$ and parameterized by $\Theta$. Lastly, the prior $p(\Theta|\mathcal{H})$ encompasses any prior knowledge we have about the given hypothesis and $p(d|\mathcal{H})$ is the marginalized likelihood or evidence
\be
\label{eq:evidence}
p(d|\mathcal{H})=\int d\Theta\,p(d|\Theta,\mathcal{H})p(\Theta|\mathcal{H}).
\ee
For the purposes of parameter estimation we can safely ignore the evidence in Bayes theorem since it is just a normalizing factor that does not depend on the model parameters $\vec\phi$. However, if we want to perform model selection to claim a detection or compare different waveforms then the evidence is crucial. In this case we can make use of the Bayesian odds ratio between models ``$A$'' and ``$B$''
\be
\mathcal{O}=\frac{p(d|\mathcal{H}_A)}{p(d|\mathcal{H}_B)}\frac{p(\mathcal{H}_A)}{p(\mathcal{H}_B)},
\ee
where the first factor is known as the Bayes Factor which is our confidence in one model over the other based on the data (henceforth we will denote the Bayes factor as $\mathcal{B}$) and the second term is the prior odds ratio for models $A$ and $B$ which describes our prior belief in both models. In this paper we consider only the Bayes factor, and assume the prior odds are even. (The choice of the prior odds will determine the false-alarm rate of a detection scheme based on the odds ratio \citep{m12}.

In this work we wish to compare the signal model $\mathcal{H}_1$ parameterized by $\{\vec\lambda, \vec\phi\}$ to the null-hypothesis model $\mathcal{H}_0$ parameterized by $\vec\phi$. When we allow the noise and GW parameters to vary simultaneously we will need to compute the evidence for both models, $\mathcal{H}_1$ and $\mathcal{H}_0$ separately. On the other hand, when we fix the noise parameters we can simply evaluate the likelihood-ratio in \eqref{eq:lnlike} and use this to compute the evidence. In this case, since the null-hypothesis model has no free parameters, the Bayes factor is simply given by the evidence computed using the likelihood ratio. In practice, to compute the evidence we make use of thermodynamic integration as detailed in E13 and Appendix \ref{sec:PTev}. The results of our Bayesian search and verification on injections will be discussed in the next section.

\subsubsection{Priors}

In a Bayesian analysis, especially when using parallel tempering and thermodynamic integration, it is very important to choose reasonable priors so that we are not exploring areas of parameter space that have been ruled out by previous experiments. We choose isotropic priors on all angular parameters and uniform priors in the log of the chirp mass with $\mathcal{M}\in[10^8,10^{10}]\,{\rm M}_{\odot}$,  luminosity distance with $d_L \in [1,10^4]$ Mpc, and frequency of the GW with $f_{\rm gw}\in [6\times 10^{-9}, 4\times 10^{-7}]$ Hz. We impose an additional condition on the average strain amplitude such that $h_0(\mathcal{M},d_L,f_{\rm gw}) \le h_{\rm ref}(f_{\rm gw}/f_0)^{2/3}$, where $h_{\rm ref}=1\times 10^{-13}$ and $f_0=10^{-8}$ Hz. This value is chosen so that the maximum strain is well above the level of detection. Essentially this is a cheap way to impose a correlated prior on chirp mass, luminosity distance, and GW frequency. The normalization is computed through Monte Carlo integration. For the pulsar distance prior we use the current electromagnetic (EM) measurements either from timing parallax or Very Long Baseline Interferometry (VLBI) corresponding to the best measured values taken from \cite{vwc+12} (10 pulsars) if available, otherwise, we use the values from the Australia National Telescope Facility (ATNF) pulsar catalog \citep{mhth05}\footnote[1]{\texttt{http://www.atnf.csiro.au/people/pulsar/psrcat/}} which have distances based on dispersion measure and the NE2001 Galactic electron-density model \citep{cl02, cl03}. For pulsars without parallax distances we assume a 20\% uncertainty on the distance. Using this information, we write the distance prior as follows
\be
p(\vec L)=\prod_{\alpha=1}^{N_{\rm psr}}\frac{1}{\sqrt{2\pi \sigma_\alpha^2}}\exp\lp -\frac{(L_{\alpha}-L^{\rm EM}_{\alpha})^2}{2\sigma_{\alpha}^2} \rp,
\ee
where $L^{\rm EM}_{\alpha}$ is the best measured distance for the $\alpha$th pulsar  and $\sigma_{\alpha}$ is the 1-sigma uncertainty on that distance measurement. In principle it would be more correct to use a Gaussian prior for the parallax, which is proportional to $L^{-1}$. If the variance on the parallax is quite large then the corresponding prior on distance will differ significantly, namely it will have a long tail towards higher distances. However, for the pulsars used in this analysis, the distance uncertainty is small enough that the two prior distributions are effectively the same and we are safe in using a  Gaussian prior on the pulsar distance itself; however, for future analyses we will move to Gaussian priors in $L^{-1}$.

For our noise parameters, we use priors that are uniform in the EFAC in the range $[0.5,5]$, uniform in the log of the EQUAD with EQUAD $\in [10^{-9}, 10^{-5}]$ s, uniform in the log of the jitter value with the same range as the EQUAD, uniform in the log of the red noise amplitude with $A_{\rm red}\in[10^{-18}, 10^{-11}]$, where the amplitude is in GW units, and uniform in the red noise spectral index with $\gamma_{\rm red}\in[1,7]$. We impose a further prior on the red noise such that the variance $\sigma_{\rm red}^2$ is less than the \emph{unweighted} standard deviation of the pulsar timing residuals, where
\be
\sigma_{\rm red}=\int_{1/T}^{\infty}df P(f) = 2.05\frac{1}{\sqrt{\gamma_{\rm red}-1}}\lp\frac{A_{\rm red}}{10^{-15}}\rp\lp \frac{T}{1\, {\rm yr}}\rp^{\frac{\gamma_{\rm red}-1}{2}}\,\rm{ns},
\ee
with $T$ the total observation time and $P(f)$ the power spectrum of the red noise.  This prior essentially restricts the model from considering red noise dominated residuals, which is a very good approximation \citep{pjl+13,esd+13}. This prior is chosen because it leads to much more computationally efficient runs by allowing us to run fewer high temperature chains in the Thermodynamic Integration scheme (See Appendix \ref{sec:PTev} for more details). In principle this red noise prior is illegal in the sense that it uses the data (i.e., the variance of the residuals); however, this prior restricts access to an area of parameter space that is not supported by the likelihood function. Therefore, by omitting this area of parameter space the evidence calculation for each model, $\mathcal{H}_1$ and $\mathcal{H}_0$, will be biased slightly low. However, the likelihood function evaluated at this area of parameter space is essentially zero, and this slight bias will be negligible.

\section{Results}
\label{sec:results}

In this section we report the results of our frequentist and Bayesian searches, provide verification of the pipeline on injected signals and report on several upper limits.

\subsection{Verification}

\begin{figure}
  \begin{center}
  \includegraphics[scale=1.0]{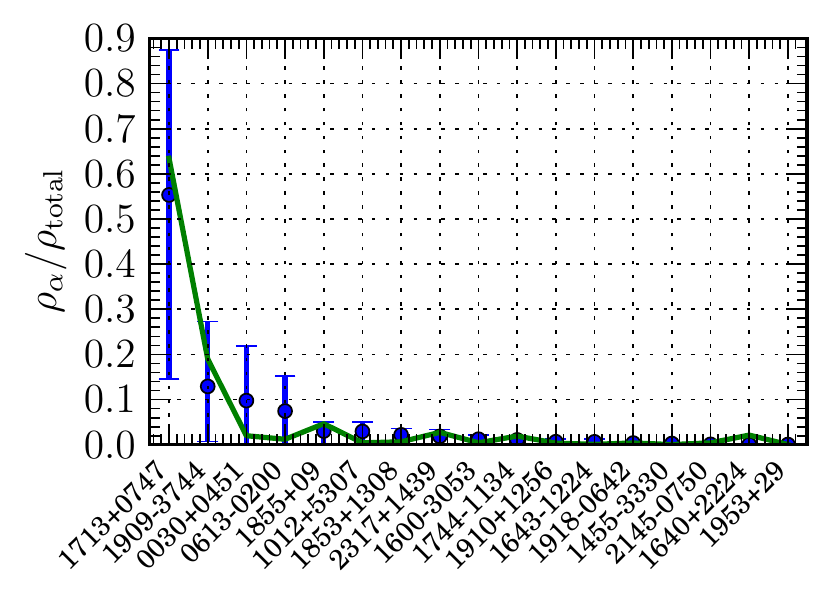}
   \end{center}
  \caption{Fraction of SNR that each pulsar contributes (black(blue) points). We see that PSR J1713$+$0747 dominates the total SNR. The gray(green) curve is a simple $1/\sigma_{\alpha}$ scaling which matches the measured SNR values quite well showing that the overall variance of the noise for each pulsar is the dominating factor in determining the overall SNR. (Color figures available in online version.)}
  \label{fig:percent_npsr}
\end{figure}

First, it is interesting to determine how much each pulsar in the 17-pulsar array will contribute to the overall SNR (signal-to-noise ratio) when a GW is present. In Figure \ref{fig:percent_npsr} we plot the fraction $\rho_{\alpha}/\rho_{\rm total}$, where $\rho_{\alpha}$ is the single pulsar SNR, for each pulsar in the array.
To compute this fraction we simulate 5000 $\rm{SNR}=10$ GW realizations (with parameters drawn from isotropic distributions in all angles and distributions uniform in the log of chirp mass and frequency) and calculate the single pulsar and total PTA SNR from Eq. \eqref{eq:snr}. The black(blue) points in the plot show the mean and standard deviation of the aforementioned ratio for each pulsar and the gray(green) curve is a simple naive scaling of $1/\sigma_{\alpha}^2$, where $\sigma_{\alpha}$ is the weighted RMS of the $\alpha$th pulsar's TOA uncertainties. It is obvious that J1713+0747 contributes more than 55\% of the SNR on average, and PSRs J1909$-$3744, J0030+0451, and J0613$-$0200 contribute $\sim$ 10\% on average. As we see from the gray(green) curve, this is very consistent with the overall scaling with the inverse of the variance of the noise; however, PSRs J0030+0451 and J0613$-$0200 carry a higher percentage because they are located opposite to the bulk of other pulsars on the sky, and therefore will contribute more to the SNR for GWs coming from that side of the sky due to the antenna pattern response. This calculation does not mean that we advocate only timing the pulsars with the highest timing precision. Although many of the lower timing precision pulsars do not help with continuous GW detection or parameter estimation, they are \emph{essential} for detection and parameter estimation of a stochastic GW background \citep[see e.g.,][]{sejr13}.

The fact that one pulsar dominates the total SNR means that it will be harder to make a confident GW detection as we require the same GW signal (with quadrupolar correlations) to be present in all pulsars. In other words, if the GW is only ``seen" in one or two pulsars then it is hard to distinguish it from some other effect due to the pulsar timing model, ISM effects or some other systematic effect. This also implies the need to run a Bayesian analysis where both the noise and GW parameters are allowed to vary simultaneously. This does not mean that a continuous GW would not be a valid interpretation of a loud sinusoidal signal in one pulsar, only that statistically we do not have enough information to confidently claim a detection. Furthermore, if we did have a loud \emph{detectable} signal, parameter estimation would be quite poor with the current NANOGrav PTA as there would be large degeneracies in the sky location (due to the small effective number of detector baselines), making sky localization and binary orientation estimates very poor. However, NANOGrav is currently timing 43 pulsars with microsecond or better precision.  Also, new ultra-wideband receivers \citep{drd+08} have increased timing precision by a factor of $\sim$1.7 for many of the pulsars in this 5-year. More pulsars and better timing precision could help ameliorate some of the limitations we have with the 5-year data set.  

Despite the potential limitations discussed above,  we verify the efficacy of our pipeline by running both the frequentist and Bayesian pipeline on a synthetic dataset with an injected GW source. To create the synthetic dataset we first compute the residuals of our 17 NANOGrav pulsars using the \textsc{tempo2} \citep{hem06} package. Next we subtract these residuals from the site arrival times, thereby producing a new set of arrival times that match our timing model perfectly. To each set of idealized TOAs we then add a Gaussian noise process with the same characteristics as those measured in the real data, and a GW signal using the fully evolving signal model. We then use these new TOAs to produce a set of synthetic residuals. For this simulation we have chosen to inject a signal with SNR 10 and parameters $\vec\lambda =\{\theta=2.07, \varphi=5.4, f_{\rm gw}=4\times 10^{-8} \rm{\,Hz}, \mathcal{M}=5\times10^{8}{\rm M}_{\odot}, \mathnormal{d_L}=1.0\rm{\,Mpc}, \psi=0.78, \iota=0.26, \Phi_0=0.53\}$.

The $\mathcal{F}_p$-statistic pipeline was run on this synthetic dataset. Since we are treating this injection as if it were a true blind search, we must first run a single-pulsar noise analysis to determine the maximum a posteriori noise parameters; however, since a strong continuous GW and red noise will be covariant we have included a single frequency sinusoid as part of our noise analysis for each pulsar. This is implemented by simply adding a free amplitude and frequency parameter to the noise model discussed above.  While this may appear to be special treatment for the injected signal, we have run the same noise model on the real data and find no evidence for any sinusoidal features. After we have obtained the maximum a-posteriori noise parameters (not including the sinusoid parameters), we use these values to construct the noise covariance matrix for use in the $\mathcal{F}_p$-statistic as well as the fixed-noise Bayesian search. By performing the noise search with an included sinusoid but not including it in our noise covariance matrix in the subsequent GW analysis we are sidestepping the problem of the GW being absorbed into red noise parameters. 

\begin{figure}
  \begin{center}
  \includegraphics[scale=1.0]{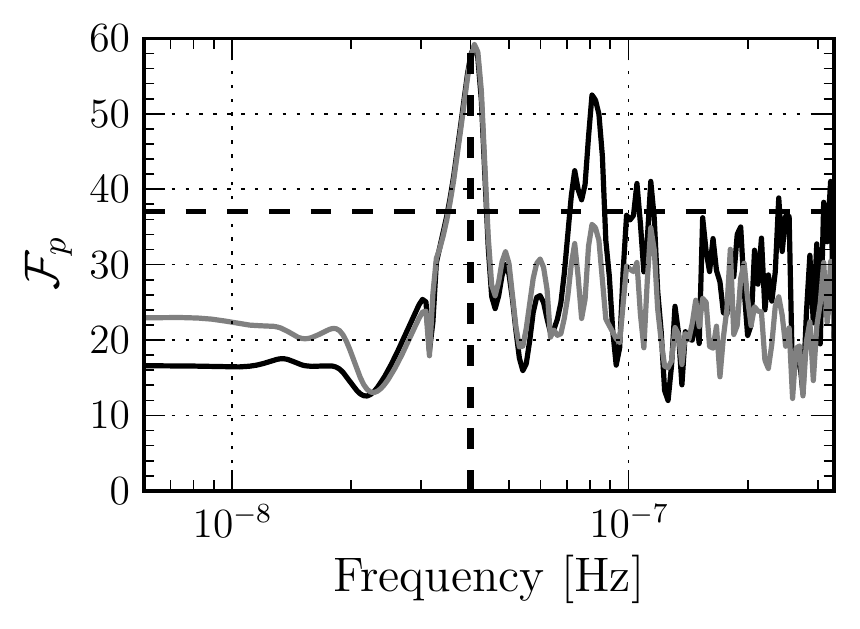}
   \end{center}
  \caption{$\mathcal{F}_p$-statistic evaluated over the frequency range $f_{\rm gw}\in[1/T,3.3\times 10^{-7}]$ Hz. The horizontal dashed line corresponds to our detection threshold of $FAP=10^{-4}$ and the vertical dashed line denotes the injected frequency. The black and gray curves are the $\mathcal{F}_p$-statistic values when using the measured and true noise parameters, respectively. See text for more details.}
  \label{fig:nano_fstat_inj}
\end{figure}

We have carried out this analysis and the results are shown in Figure \ref{fig:nano_fstat_inj} where we plot $\mathcal{F}_p$ vs. GW frequency when using the measured noise values (black) and the true injected noise values (grey). The vertical dashed line indicates the injected frequency and the horizontal dashed line represents our detection threshold corresponding to a FAP of $10^{-4}$. To compute the total FAP over all frequencies we make use of Eq. \eqref{eq:FAPT} and choose $N_f = 324$, resulting in a total FAP of $1.6\times10^{-8}$ which is a decisive detection. The number of independent frequencies is difficult to calculate when we are using many datasets with very irregular sampling. In this work we have chosen $N_f=324$ as this corresponds to $f_{\rm gw}\in[1/T,3.3\times 10^{-7}]$ and $\Delta f_{\rm gw} = 10^{-9}$ Hz. The upper limit on frequency was chosen because our approximate observing cadence is 2.5 weeks$^{-1}$ and the frequency spacing was chosen by imposing the condition that the autocorrelation function of $\mathcal{F}_p$ when no signal is present drops to half of its maximum value at that frequency lag. This analysis shows us that we can indeed detect a continuous GW if it is present in our data by conducting a fully blind search; however, we also see that our results will not be conclusive as there are several frequencies at which the $\rm{FAP}$ is above our threshold value. From comparison with true-noise case, we see that the uncertainty (and residual correlations between GW and noise parameters) in the noise parameters can lead to confusing results. This again, is mostly due to the fact that our sensitivity is dominated by a small number of pulsars. Because of this, we caution against using a fixed-noise method to make final detection statements but instead advocate these methods as a first round in a suite of analyses.

Both Bayesian pipelines (with and without varying noise parameters) were run on this synthetic dataset. For both runs we have used PTMCMC and thermodynamic integration as discussed in Appendix \ref{sec:PTev}. Due to the large parameter spaces when using the full GW and noise model, we have chosen to use only the pulsars that contribute more than 1\% to the injected SNR, resulting in 6 pulsars, J1713$+$0747, J1909$-$3744, B1855$+$09, J0030$+$0451, J0613$-$0200, and J1012$+$5307. Here we use the same noise parameters as mentioned above for our fixed-noise search. Even though these estimates are different from the true noise parameters, we nonetheless achieve a log-Bayes factor of 27.4 for the fixed-noise search (a log-Bayes factor greater than 5 is considered decisive evidence). However, as we mentioned earlier, we should not totally trust this level of evidence as it does not fully incorporate our uncertainty in the noise model. When we run an analysis where we allow the noise and GW parameters to vary simultaneously we only achieve a log-Bayes factor of 5.35. While still decisive, this search is much less sensitive to the GW; nonetheless, this search is the most robust and will be the real test as to whether or not one can trust a real GW detection candidate. Of course these results could change depending on the noise realization or GW parameter combinations. A more detailed study of this is warranted but beyond the scope of this paper. Nonetheless, the large spread of overall noise levels in modern PTAs will most likely make the confident detection of a continuous wave GW very difficult.

\subsection{Search Results}
\begin{figure}
  \begin{center}
  \includegraphics[scale=1.0]{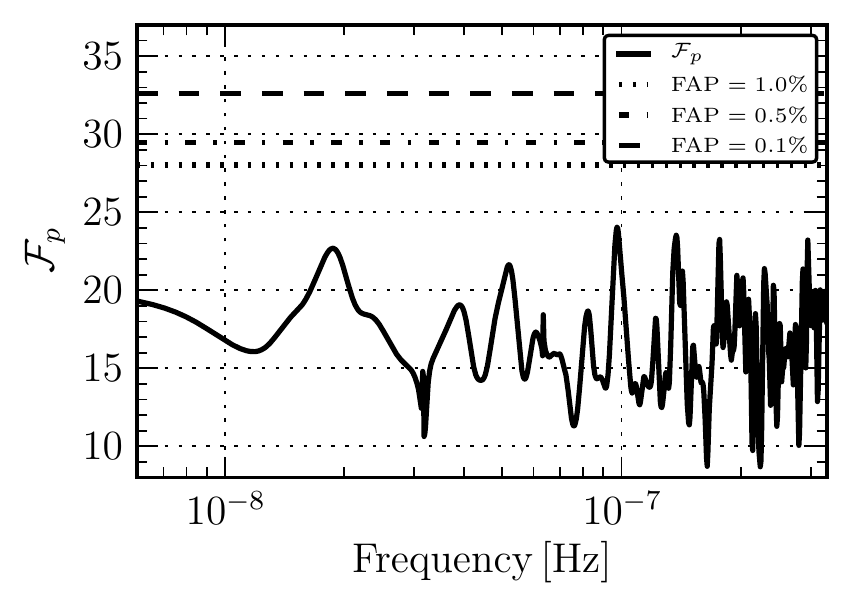}
   \end{center}
  \caption{$\mathcal{F}_p$-statistic evaluated over the frequency range $f_{\rm gw}\in[1/T,3.3\times 10^{-7}]$ Hz.  These frequencies were chosen based on the fact that the approximate cadence is 2.5 weeks$^{-1}$. The dashed, dash-dotted, and dotted lines represent the value of $\mathcal{F}_p$ that gives a FAP of 0.1\%, 0.5\%, and 1\%, respectively. Here we note that there is no evidence for a detection and the data are consistent with the null hypothesis.}
  \label{fig:nano_fstat}
\end{figure}

First we will discuss the results of the $\mathcal{F}_p$-statistic search on the real 17-pulsar NANOGrav data. To carry out the analysis we have computed $\mathcal{F}_p$ for many frequencies with $f_{\rm gw}\in[1/T,3.3\times 10^{-7}]$ Hz. These frequencies were chosen based on the fact that the approximate cadence is 2.5 weeks$^{-1}$. The results of this search are shown in Figure \ref{fig:nano_fstat} where the solid black line is the value $\mathcal{F}_p$ at each frequency, the dotted, dash-dotted, and dashed lines are the value of $\mathcal{F}_p$ corresponding to a 1.0\%, 0.5\%, and 0.1\% FAP, respectively, where these values are calculated from Eq. \eqref{eq:FAP}. Furthermore if we maximize $\mathcal{F}_p$ over frequencies then the total FAP, accounting for the trials factor $N_f$ is very nearly 1, indicating that we should accept the null hypothesis (no visible GW signal) with very high confidence.

We will now briefly discuss the results of both Bayesian searches. To carry out this analysis we have run our PTMCMC and computed the Bayes factors for each case. In the first case we allow the noise parameters and GW parameters to vary and explicitly compute the Bayesian evidence via thermodynamic integration for a model with a GW and noise and a model with just noise. In the second case, we fix the noise parameters to the maximum a-posteriori obtained from single pulsar analyses and only allow the GW parameters to vary. As mentioned above, the second case is not reliable since there is likely to be correlations between the GW and noise parameters; however, we give the results of both searches for completeness. As above, in the case of a true continuous GW signal we can get very different results from a fixed-noise versus a varying noise search. However, in our case the log-Bayes factor for searches with and without varying noise parameters is $-0.55$ and $-0.1$, respectively, both indicating that there is no evidence for a continuous GW and a model consisting of noise is preferred. We further note that this is completely consistent with our frequentist analysis.

\subsection{Upper Limits}

In this section we will outline the procedures used to compute both the frequentist and Bayesian upper limits on the strain amplitude, $h_0$. First we wish to make state that an $x\%$ upper limit on the strain amplitude does not mean that we would have detected a signal with that amplitude with 95\% confidence, it simply means that the \emph{true} value of the amplitude is less than the upper limit with 95\% probability (this probability measures the frequency of measuring that value of the amplitude in the frequentist case and the degree of belief in the true value of the amplitude in the Bayesian case).  In the following sections we will discuss the mathematics of upper limit computation in the frequentist and Bayesian frameworks, and then we will lay out our computational procedure.

\subsubsection{Frequentist Approach}

From a frequentist viewpoint, the data are random while the signal parameters are fixed but unknown (i.e., we construct probability distributions for the data, or rather some function of the data, given a set of signal parameters), whereas in the Bayesian framework the data are fixed and the signal parameters are uncertain (i.e., we construct probability distributions of the signal parameters given a dataset). From the above statement it then follows that frequentist upper limits are derived from integrating the probability distribution of some statistic of the data (the $\mathcal{F}_p$-statistic in this case) at a fixed value of the parameter of interest, and Bayesian upper limits are derived from integrating the probability distribution of the parameter of interest for the given data set. 

More formally, the probability distribution of the $\mathcal{F}_p$-statistic given a value of the strain amplitude $h_0$ is
\be
p(\mathcal{F}_p|h_0)=\int p(\mathcal{F}_p|h_0,\tilde{\lambda},\mathbf{n})p(\tilde{\lambda})p(\mathbf{n})\,d\tilde\lambda \,d\mathbf{n},
\ee
where $\tilde\lambda=\{\theta,\varphi, f_{\rm gw}, \mathcal{M}, \iota, \psi, \Phi_0\}$ is a reduced parameter space vector, $p(\tilde\lambda)$ is the sampling distribution of $\tilde\lambda$ (these sampling distributions are identical to the prior probability distributions in the bayesian case), $\mathbf{n}$ is a noise timeseries drawn from the distribution
\be
p(\mathbf{n})=\frac{1}{\sqrt{\det{2\pi C}}}\exp{\lp -\frac{1}{2}\mathbf{n}^TC^{-1}\mathbf{n} \rp},
\ee
with $C$ the covariance matrix of the noise in the pulsar timing residual timeseries, and $p(\mathcal{F}_p|h_0,\tilde\lambda,\mathbf{n})$ is the probability distribution function for the $\mathcal{F}_p$ statistic for given values of $h_0$ and $\tilde\lambda$ and a given noise realization $\mathbf{n}$. An upper limit on $h_0$ at confidence level $\alpha$ is then computed as 
\be
\begin{split}
\label{eq:freq_upper}
\alpha &= \int_{\mathcal{F}_{p,0}}^{\infty} p(\mathcal{F}_p|h_0)\,d\mathcal{F}_p\\
&= \bigg\langle\frac{1}{N} \sum_{i=1}^{N}
\begin{cases}
1 & {\rm if\,\,} \mathcal{F}_{p,i} \ge \mathcal{F}_{p,0}\\
0 & {\rm otherwise}
\end{cases}\bigg\}
\bigg\rangle,
\end{split}
\ee
where the $N$ observables $\mathcal{F}_{p,i}$ are drawn from the ``signal distribution'', $p(\mathcal{F}_p|h_0)$, and the average, $\langle \cdot \rangle$, is over that distribution. In other words, we integrate the probability distribution of the $\mathcal{F}_p$-statistic over the so called ``signal space'' (i.e., from the measured value $\mathcal{F}_{p,0}$ to infinity), that is, we count the number of signal realizations that gives an $\mathcal{F}_p$-statistic value larger than the one measured in the actual dataset. This integral can take on any value $\alpha \in [0,1]$ for a given $h_0$; therefore, the integral must be repeated with different values of $h_0$ until $\alpha = 0.95$ for a $95\%$ upper limit.

In practice, we carry out the following computational procedure:
\begin{enumerate}

\item Measure the value $\mathcal{F}_{p,0}$ from the real 17-pulsar NANOGrav dataset as described in Section \ref{sec:fpstat}.

\item Simulate a synthetic noise vector $\mathbf{n}=L\mathbf{w}$ for each pulsar, where $L$ is the Cholesky decomposition of the noise covariance matrix $C$, and $\mathbf{w}$ is a unit variance, zero-mean, vector.

\item Choose strain amplitude, $h_0$ and construct a GW waveform $\mathbf{s}(t,h_0,\tilde\lambda)$ for each pulsar where the parameters, $\tilde\lambda$ are drawn from the distribution $p(\tilde\lambda)$.

\item Construct a new set of residuals for each pulsar $\delta t_{\rm sim} = R\lp\mathbf{n} + \mathbf{s}(t, h_0,\tilde\lambda)\rp$, where $R$ is the so called fitting projection matrix introduced in \cite{dfg+12} and \cite{ esvh13}.\footnote{We choose to create residuals with the $R$ matrix rather than re-fitting the timing model with \textsc{tempo2} in order to simulate many datasets quickly. We have done many tests to make sure that we get the same results using both the $R$ matrix and using a full \textsc{tempo2} run.} 

\item Now measure the value $\mathcal{F}_{p,i}$ for the simulated dataset.

\item Repeat steps 2--5 10,000 times and measure the number of realizations that result in $\mathcal{F}_{p,i} > \mathcal{F}_{p,0}$.

\item Repeat steps 2--6 with different values of $h_0$ until 95\% of simulations result in $\mathcal{F}_{p,i} > \mathcal{F}_{p,0}$.

\end{enumerate}
In the remainder of the paper we will choose to compute upper limits on the strain amplitude as a function of GW frequency or GW sky location at a fixed GW frequency. To facilitate such upper limits we simply fix the parameters (either GW frequency or sky location) when simulating waveforms in step 3.

\subsubsection{Bayesian Approach}
As mentioned above, in the Bayesian framework we do not rely on simulations as we treat the data as fixed and integrate the posterior pdf of the parameter of interest to compute the upper limit. In principle, a Bayesian upper limit is much more simple and intuitive than a frequentist upper limit. To compute a Bayesian upper limit we compute an integral that is analogous to Eq. \eqref{eq:freq_upper}
\be
\begin{split}
\alpha &= \int_0^{h_{\rm up}}dh_0 \int d\tilde\lambda\, d\vec\phi\, p(\delta t|h_0,\tilde\lambda,\vec\phi)p(h_0)p(\tilde\lambda)p(\vec\phi)\\
&= \int_0^{h_{\rm up}}dh_0\,p(\delta t|h_0)p(h_0),
\end{split}
\ee
where $p(\delta t|h_0,\tilde\lambda,\vec\phi)$ is the likelihood function, $p(h_0)$, $p(\tilde\lambda)$, $p(\vec\phi)$ are the prior probability distributions on $h_0$, $\tilde\lambda$, and $\vec\phi$, respectively, where $\vec\phi$ denotes the noise model parameters. In words, we simply integrate the marginalized posterior distribution of $h_0$ until the desired credible region corresponding to a probability of $\alpha$ is reached at $h=h_{\rm up}$. As in the frequentist case, we want upper limits on the strain amplitude as a function of GW frequency or sky location. In this case we simply fix the parameters and then marginalize over the others. In practice, to compute the Bayesian upper limits we carry out a separate MCMC run for fixed values of frequency and/or fixed sky locations and then compute the 95\% upper limit for each. The choice of prior on $h_0$ is very important and can lead to very different upper limits. Such a detailed analysis of priors is beyond the scope of this work but will be addressed in a future paper. In principle, our prior distribution should come from population synthesis models \citep{s13}; however, since we wish our upper limits to be informed by our data and not dominated by our prior distribution we use a very conservative\footnote{On a logarithmic scale this prior prefers higher strain values a priori; however, it is conservative in the sense that the corresponding upper limit will not overestimate our sensitivity and the limit will not depend on the lower bound of the prior as is the case for logarithmic priors.} prior that is uniform in $h_0$ with $h_0 \in [0,10^{-11}]$.

\subsubsection{Sky Averaged Strain Upper Limits}

\begin{figure*}
  \begin{center}
  \includegraphics[scale=1.0]{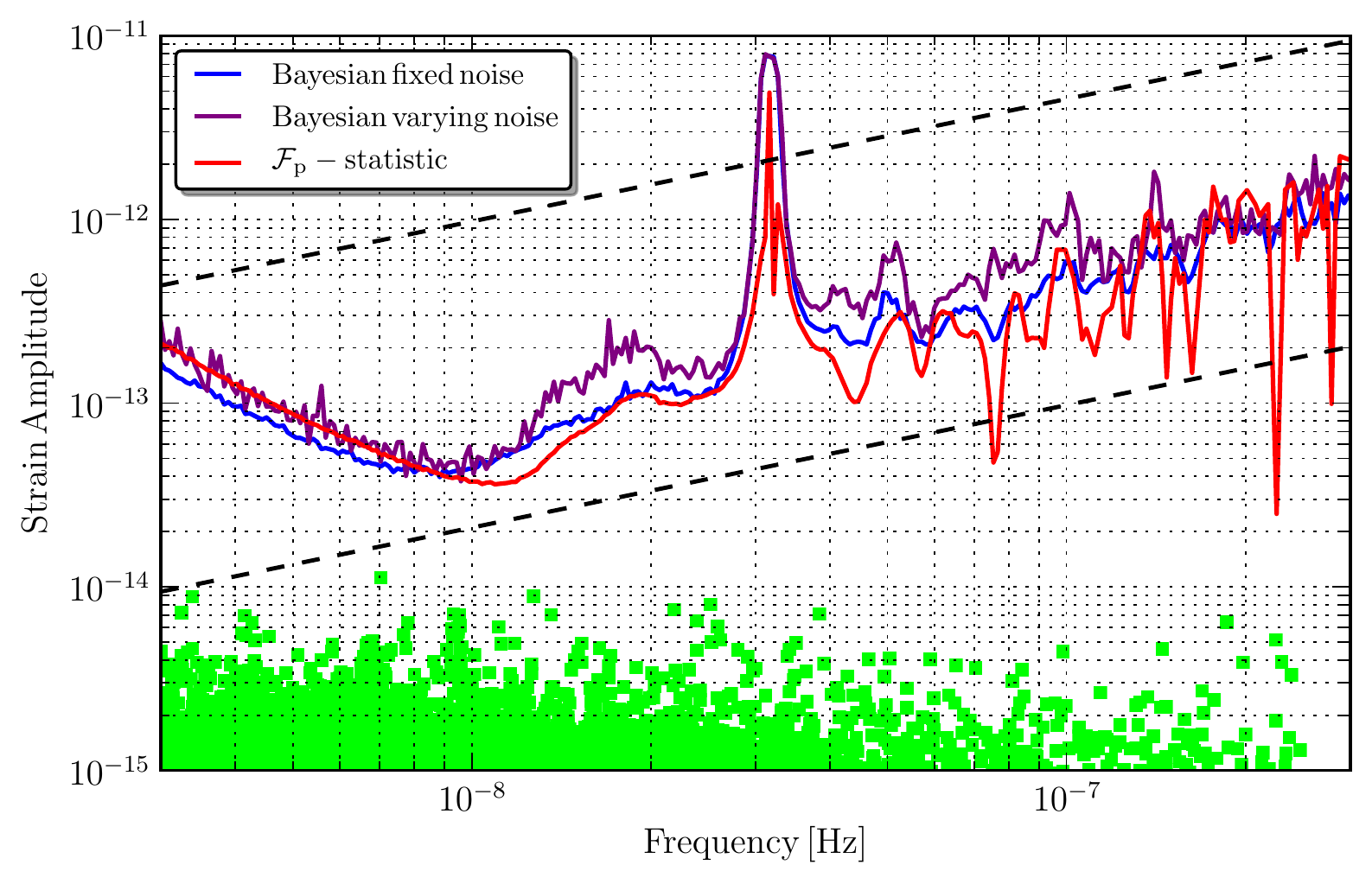}
   \end{center}
  \caption{Sky-averaged upper limit on the strain amplitude, $h_0$ as a function of GW frequency. The Bayesian upper limits are computed using a fixed-noise model (thick black(blue)) and a varying noise model (thin black(purple)) and the frequentist upper limit (gray(red)) is computed using the $\mathcal{F}_p$-statistic. The dashed curves indicate lines of constant chirp mass for a source with a distance to the Virgo cluster (16.5 Mpc) and chirp mass of $10^{9}{\rm M}_{\odot}$ (lower) and $10^{10}{\rm M}_{\odot}$ (upper). The gray(green) squares show the strain amplitude of the loudest GW sources in 1000 monte-carlo realizations using an optimistic phenomenological model of \cite{s13}. See text for more details. (Color figures available in online version.)}
  \label{fig:upper}
\end{figure*}
In Figure \ref{fig:upper} we report the 95\% upper limits on the strain amplitude, $h_0$, as a function of GW frequency computed using the methods described above for the frequentist and Bayesian pipelines. The gray(red), thick black(blue) and thin black(purple) curves are the 95\% upper limits on strain amplitude computed using the $\mathcal{F}_p$-statistic, Bayesian method with fixed-noise values, and Bayesian method with varying noise values, respectively. There are several features in the plot that require explanation. First, the decrease in sensitivity at $f_{\rm gw}=1\rm{yr}^{-1}$ and $f_{\rm gw}=2\rm{yr}^{-1}$ is due to the sky position  and parallax fitting in the timing model, respectively. The upward trend at lower frequencies is due to the quadratic spin-down model fit. The noisiness of the frequentist upper limit is due to the fact that $\mathcal{F}_p$-statistic distribution at higher frequencies is indeed quite noisy when computed using the real data, and since our upper limits compare the value measured in real data to values measured in simulated data, this noisiness is to be expected.

If we compare our results to those of \cite{yhj+10}, we see that the upper limits using the 5-year NANOGrav datasets are a factor of 2 to 3 times more constraining. The main reason for this improvement is the higher timing precision of the NANOGrav dataset as compared to the older PPTA data sets \citep{vbc+09}. Although the procedures for setting frequentist and Bayesian upper limits is quite different, our results are very similar. In part, this is due to the fact that we have used a uniform prior on the strain amplitude, $h_0$, making our Bayesian analysis very similar to a pure likelihood analysis. Since the $\mathcal{F}_p$-statistic is just the likelihood (ratio) maximized over amplitudes, we would expect a likelihood analysis to give similar results. Note that the Bayesian upper limits when varying the noise parameters are somewhat less constraining than the fixed-noise case. This is to be expected since at lower frequencies the GW amplitude and red noise amplitude are somewhat correlated and at higher frequencies the  GW amplitude and jitter parameter are somewhat correlated. Both correlations will result in slightly worse upper limits on the GW amplitude when allowing the noise parameters to vary.

In Figure \ref{fig:upper}, the dashed curves indicate lines of constant chirp mass for a source with a distance to the Virgo cluster (16.5 Mpc) and chirp mass of $10^{9}$ and $10^{10}$, respectively and the gray(green) squares are the strain amplitude of the loudest GW events in 1000 Monte Carlo realizations using an optimistic phenomenological model of \cite{s13}. The model used here produces a stochastic GW background with dimensionless strain amplitude of $\sim 2\times 10^{-15}$, just below the current upper limits presented in \cite{src+13}. astrophysically, these upper limits tell us that we can essentially rule out any source with $\mathcal{M}\ge 10^{10}{\rm M}_{\odot}$ at the distance to the Virgo cluster (16.5 Mpc); however, our horizon distance falls just short of the Virgo cluster for sources with $\mathcal{M}\le 10^9{\rm M}_{\odot}$. Furthermore, we see that all sources from monte-carlo realizations (gray(green) squares) have strain amplitudes below our upper limits indicating that it is very unlikely that we will see a resolvable source at the current sensitivity (consistent with our search results). It is important to note; however, that these strain amplitude upper limits are averaged over sky location and inclination angle (either through marginalization in the Bayesian case, or from Monte Carlo sampling in the frequentist case), both of which play a large part in the \emph{overall} amplitude of the signal. Therefore, these results have the caveat that they make statements about the \emph{average} sensitivity to such GW sources; however, it is still unlikely (i.e., probability of detection $\lesssim 50\%$) that we could detect even the loudest optimally oriented source shown in Figure \ref{fig:upper}. For face on systems (i.e., $\iota=\pi/2$) and sky location near the best timed pulsars, the overall amplitude of the GW can be $\sim$ 5 times larger than the averaged strain amplitudes reported here.

\subsubsection{Angular Upper Limits}

\begin{figure*}
  \begin{center}
  \includegraphics[scale=1.0]{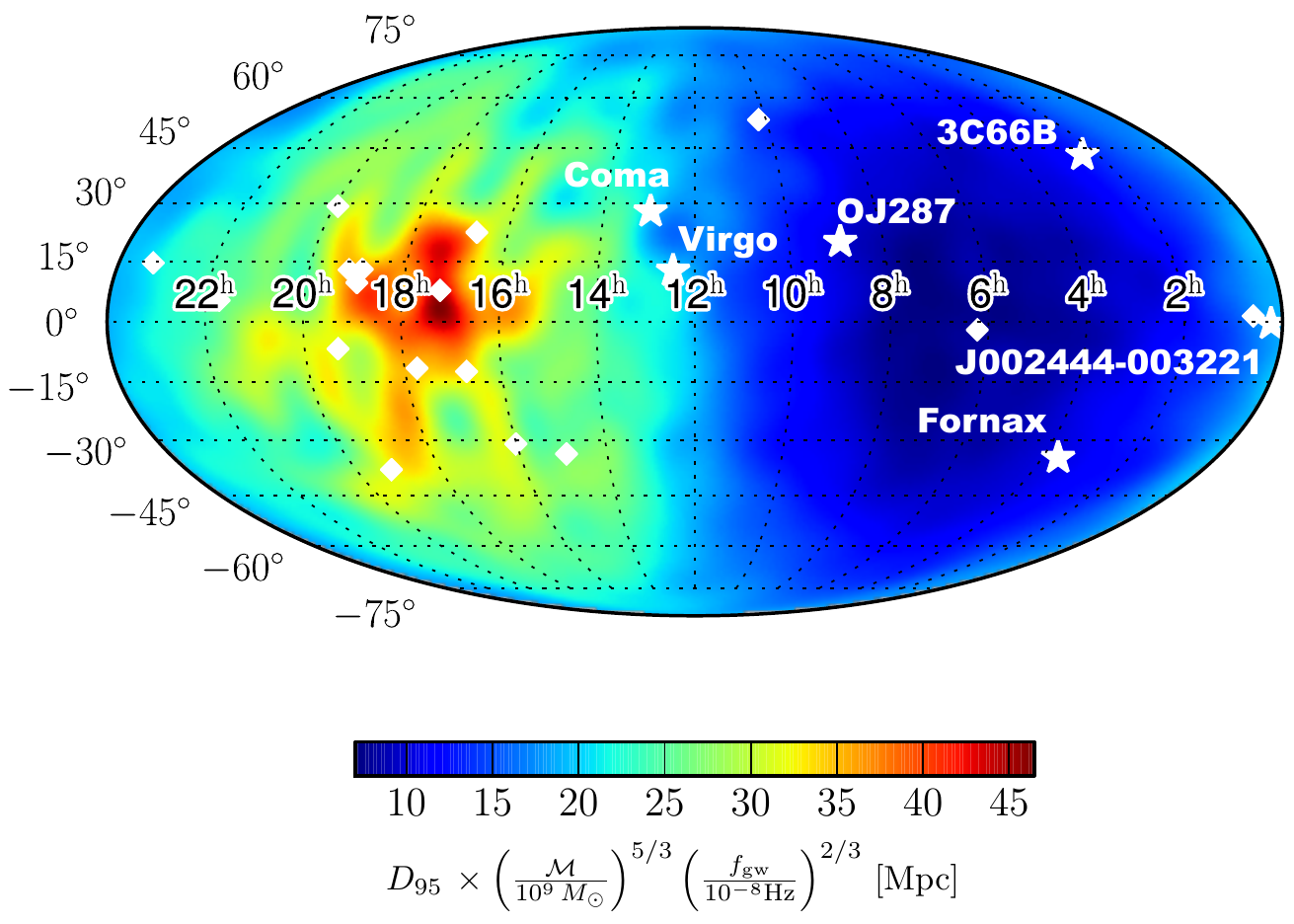}
   \end{center}
  \caption{95\% lower limit on the luminosity distance as a function of sky location computed using the $\mathcal{F}_p$-statistic plotted in equatorial coordinates . The values in the colorbar are calculated assuming a chirp mass of $\mathcal{M}=10^9{\rm M}_{\odot}$ and a GW frequency $f_{\rm gw}=1\times 10^{-8}$ Hz. The white diamonds denote the locations of the pulsars in the sky and the black(white) stars denote possible SMBHBs or clusters possibly containing SMBHBs. As expected from the antenna pattern functions of the pulsars, we are most sensitive to GWs from sky locations near the pulsars. The luminosity distances to the potential sources are 92.3, 1575.5, 2161.7, 16.5, 104.5, and 19 Mpc for 3C66B, OJ287, J002444$-$003221, Virgo Cluster, Coma Cluster, and Fornax Cluster, respectively.  (Color figure available in the online version.)}
  \label{fig:upper_sky}
\end{figure*}

\begin{figure*}
  \begin{center}
  \includegraphics[scale=1.0]{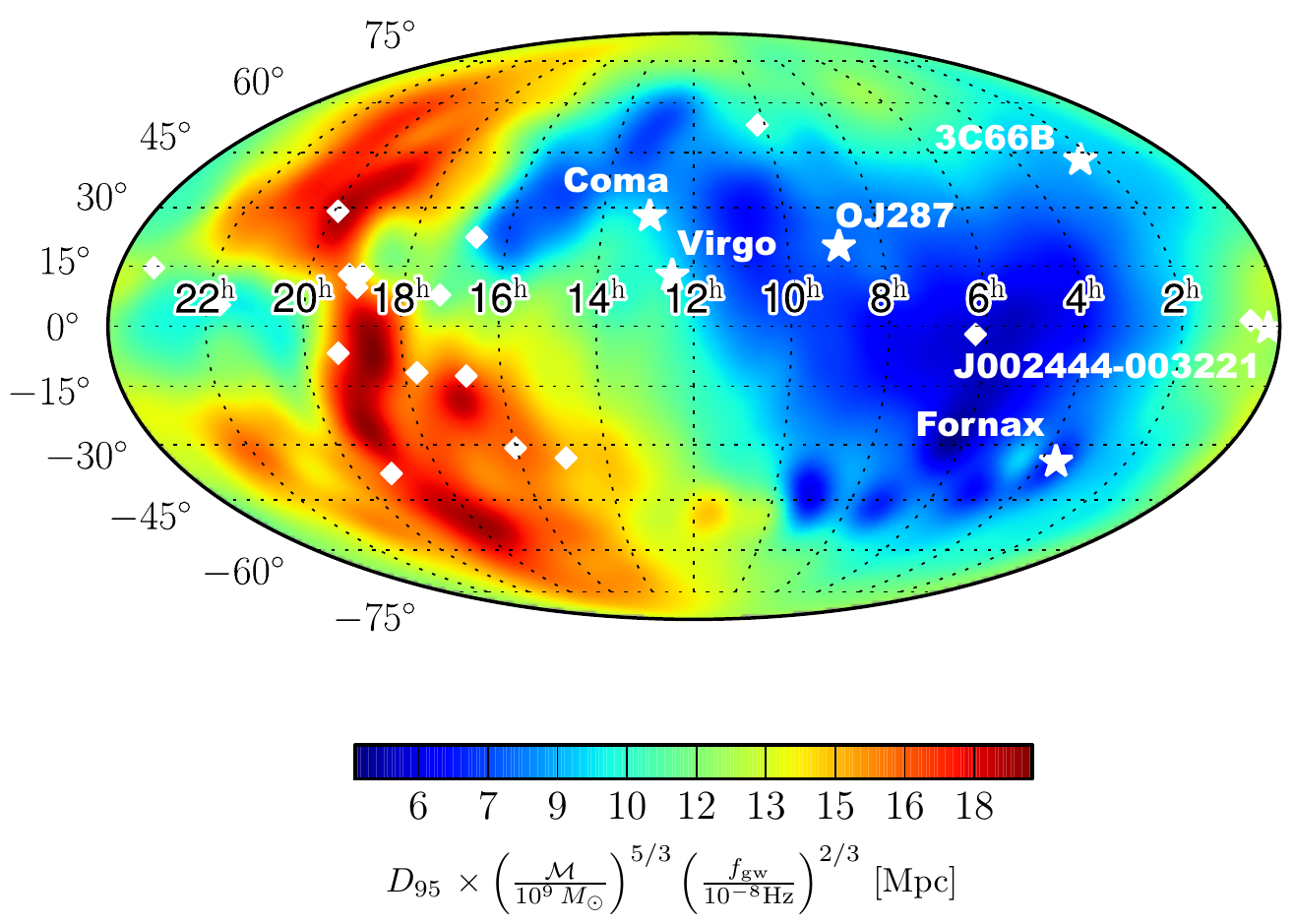}
   \end{center}
  \caption{ 95\% lower limit on the luminosity distance as a function of sky location computed using the Bayesian method including the noise model. The values in the colorbar are calculated assuming a chirp mass of $\mathcal{M}=10^9{\rm M}_{\odot}$ and a GW frequency $f_{\rm gw}=1\times 10^{-8}$ Hz. The white diamonds denote the locations of the pulsars in the sky and the black(white) stars denote possible SMBHBs or clusters possibly containing SMBHBs. (Color figure available in the online version.)}
  \label{fig:upper_sky_bayes}
\end{figure*}

In Figures \ref{fig:upper_sky}  and \ref{fig:upper_sky_bayes} we report the 95\% \emph{lower} limit on the luminosity distance as a function of sky location computed using the frequentist and Bayesian techniques, respectively. We have chosen to present our results in terms of the luminosity distance instead of the strain amplitude as it is a true \emph{physical} parameter and it gives a more intuitive feel as to what the data can constrain. To compute this lower limit we carry out the same procedure as above but we fix the frequency to $f_{\rm gw}=10^{-8}$ Hz and compute an upper limit on the strain amplitude as a function of sky location; we can then use Eq. \eqref{eq:strain_amp} to convert an upper limit on strain amplitude into a lower limit on luminosity distance. 
The values in the color bar are calculated assuming a chirp mass of $\mathcal{M}=10^{9}{\rm M}_{\odot}$ and a frequency of $f_{\rm gw}=10^{-8}$ Hz but this can be scaled to determine the minimum luminosity distance for any chirp mass value and GW frequency. In Figures \ref{fig:upper_sky} and \ref{fig:upper_sky_bayes} the white diamonds represent the locations of the 17 NANOGrav pulsars used in the analysis and the black(white) stars are the sky locations of potential GW hotspots \citep{sls+13} and possible GW source candidates \citep{vln+08,ios10, jgr+13}. 

We will now discuss the features of this sky-dependent upper limit computed using the frequentist $\mathcal{F}_p$-statistic. Firstly, we notice that the overall distribution is quite similar to the antenna pattern response (i.e., $1+\cos\mu$) as is to be expected in the case of no detection. Due to this, we are most sensitive (larger lower limit on luminosity distance) at sky locations near the best timed pulsars (i.e., J1713+0747, B1855+09, J1909-3744) and least sensitive in the opposite direction. More quantitatively, we note that in the most sensitive areas of the sky we can constrain  the luminosity distance $d_L\gtrsim 47$ Mpc for $\mathcal{M}=10^{9} {\rm M}_{\odot}$. Furthermore, it is possible to constrain the luminosity distance $d_L\gtrsim \sim 2$ Gpc in the most sensitive sky locations if we consider $10^{10}{\rm M}_{\odot}$ chirp mass sources. It should be noted that the Bayesian fixed-noise search gives nearly identical results to the fixed-noise frequentist search.

We now move to the sky-dependent upper limit computed using the full Bayesian technique where the GW and noise parameters are varied simultaneously. The first observation that we make is that the overall scale is about a factor of 2 lower than the fixed-noise frequentist or Bayesian upper limit. At first this may be surprising given the general agreement of the sky-averaged upper limits of Figure \ref{fig:upper}; however, full Bayesian sky-dependent upper limits exacerbate the problem of relatively few pulsars contributing to the overall PTA sensitivity as shown in Figure \ref{fig:percent_npsr}. Another difference in this upper limit, as opposed to the frequentist upper limit, is that it does not quite match the expected antenna pattern response function. These differences are due to the fact that we are simultaneously varying the GW and noise parameters, and when only one or a few pulsars contribute to the PTA sensitivity, there is a degeneracy between intrinsic red-noise processes in the pulsar and a common GW among all pulsars. In other words, it is very difficult to distinguish between a low-frequency continuous GW and a red noise process if only a small number of pulsars have sufficiently low noise levels to resolve the GW. 

Because Bayesian upper limits marginalize or integrate over all parameters except the amplitude, the correlations between the GW and the red noise amplitude will broaden the 1-d pdf of the amplitude and thus will result in larger upper limits as opposed to the fixed-noise case. As is clear from Figure \ref{fig:upper_sky_bayes}, the aforementioned effect is very strong for GW sky locations near our best timed pulsars. For example, we are not most sensitive to GWs around the sky location of PSR J1713$+$0747 because this pulsar contributes a very large percentage of the overall SNR of the GW in this case and thus results in a very large correlation between the GW and red noise amplitudes. 

Since, at the moment, we have no way of measuring the noise properties of the pulsars independently of any GWs that may be present in the data, to perform a completely robust upper limit or search we must allow both to vary simultaneously. Given this reality, we must view any fixed-noise results with the caveat that they assume that the noise parameters are measured perfectly and are independent of any GWs in the data. 

Unfortunately, many of the GW hotspots and potential SMBHB sources are located at insensitive sky locations, for both frequentist and Bayesian analyses, where our lower limit on distance only allows us to constrain $10^{10}{\rm M}_{\odot}$ sources. This fact is a great argument for aggressive pulsar search campaigns and the addition of new pulsars to the PTA at sky locations that are currently insensitive \citep{blf11}. 



\subsubsection{Constraints on  the SMBHB Coalescence Rate}

A non-detection of continuous GW, as we have presented here, allows us to compute an upper limit on the rate of SMBH coalescences using methods presented in \cite{wjy+11}. Since we have made no detections, we assume Poisson statistics for the probability of an event (i.e., a detectable signal) occurring, that is, the probability of no events is $e^{-\langle N\rangle}$, where $\langle N\rangle$ is the expected number of events. We use this probability distribution function to place a  95\% upper limit on the expected number of events such that $\exp(-N_{95})=0.05$, telling us that $\langle N \rangle \le \langle N_{95} \rangle = 3$. Therefore, if the expected number of events were greater than 3, at least one source would have been detected with 95\% probability. Now, following \cite{wjy+11}, the expected number of events is
\be
\begin{split}
\langle N \rangle &= \int \frac{d^2R}{d\log_{10}(1+z)\log_{10}(\mathcal{M}_r)} \lp \frac{df}{dt} \rp^{-1}\\
& \times P_d(\mathcal{M}_r,z,f)\, d\log_{10}(1+z)\, d\log_{10}(\mathcal{M}_r)\,df,
\end{split}
\ee
where $P_d(\mathcal{M}_r,z,f)$ is the probability of detecting an SMBHB with chirp mass $\mathcal{M} = \mathcal{M}_r(1+z)$, redshift $z$, and observed GW frequency $f$.  Following the derivation in \cite{wjy+11} and making the assumption that the differential coalescence rate does not vary significantly over the range $\Delta\log_{10}{\mathcal{M}_r}=1$ and $\Delta\log_{10}(1+z)=0.2$, it is possible to show that
\be
\frac{d^2R}{d\log_{10}(1+z)\log_{10}(\mathcal{M}_r)} < \frac{15}{\int \lp \frac{df}{dt} \rp^{-1}P_d(\mathcal{M}_r,z,f)\, df}.
\label{eq:coal_rate}
\ee

In order to compute the detection probability $P_d$, we make use of the $\mathcal{F}_p$ statistic. We use the same method that we have used for the upper limits, except now we compare the the value of $\mathcal{F}_p$ computed using simulated data with injections to a specified threshold based on a FAP of $10^{-4}$. We use 10,000 realizations at each value of $z$ and $f$. After the probability of detection is computed, we numerically integrate the above expression to obtain a limit on the differential coalescence rate. It should be noted that we will be able to place more meaningful constraints on the coalescence rate using upper limits on the amplitude of a stochastic background of SMBHBs; however, this is beyond the scope of this work and will be addressed in a future paper.
\begin{figure}
  \begin{center}
  \includegraphics[scale=1.0]{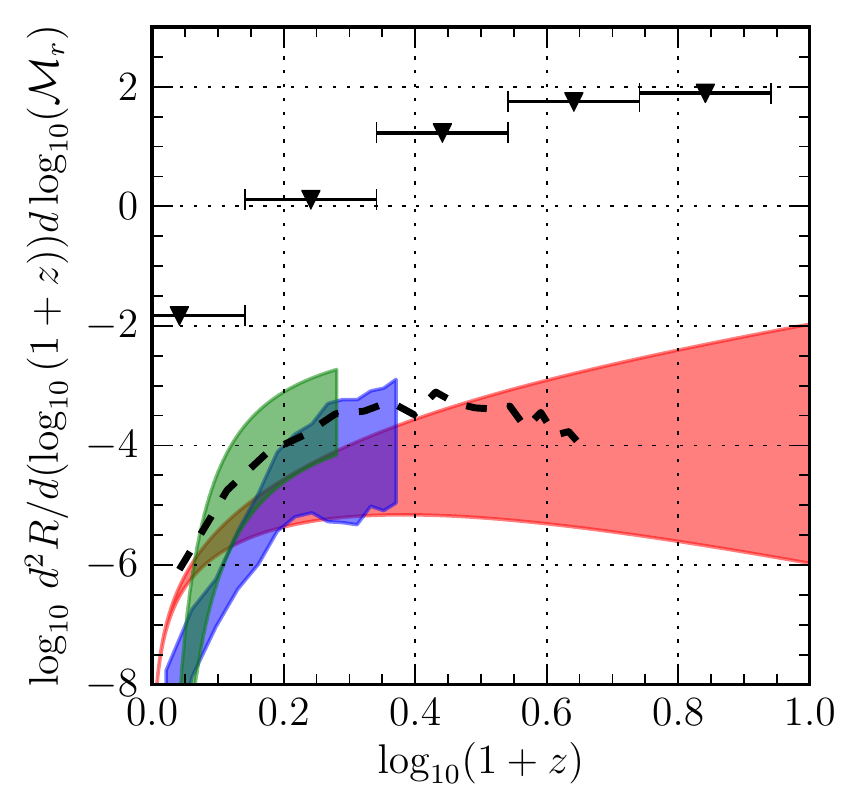}
   \end{center}
  \caption{Differential coalescence rate of SMBHBs per redshift per chirp mass with mass bin centered on $10^{10}{\rm M}_{\odot}$ and width 1 dex. We have chosen to explore only the highest masses since these high mass sources are the ones likely to be detected by GW searches in the future. The black triangles represent our upper 95\% upper limits, the light gray(red) shaded area show expected coalescence rate estimates obtained from \cite{jb03} as well as data from the Sloan Digital Sky Survey \citep{wlh09}. The medium gray(blue) shaded region comes from the phenomenological models of \cite{s13} and the black dashed line comes from an a posteriori implementation of the \cite{mm13} MBH-sigma relation into the semi-analytic model of \cite{gwb+11}. The black(green) shaded region is constructed by using the observed evolution of the galaxy mass function combined with the MBH-M-stars relation from \cite{mm13} to calibrate an analytical model for evolving the mass function via mergers \citep{mop13}.}
  \label{fig:coal_rate}
\end{figure}
In Figure \ref{fig:coal_rate} we plot our constraints on the differential coalescence rate as a function of redshift. Since we have made the assumption that this differential coalescence rate does not vary significantly over an order of magnitude in chirp mass, the results presented here are for the $\mathcal{M}_r=10^{10}{\rm M}_{\odot}$ case. We are unable to place meaningful constraints on less massive systems. The light gray(red) shaded area is constructed using the model presented in \cite{jb03} along with measurements from the Sloan Digital Sky Survey \citep{wlh09}. The medium gray(blue) shaded area is constructed by considering the different galaxy merger rates based on observations \citep{s13} along with the most recent MBH-sigma relation from \cite{mm13}. The dashed line comes from an a posteriori implementation of the \cite{mm13} MBH-sigma relation into the semi-analytic model of \cite{gwb+11} assuming accretion onto both SMBHs before merger. The black(green) shaded region is constructed by using the observed evolution of the galaxy mass function combined with the MBH-M-stars relation from \cite{mm13} to calibrate an analytical model for evolving the mass function via mergers \citep{mop13}. The figure shows that the coalescence rate for MBHs of $\sim10^{10} {\rm M}_{\odot}$  is poorly constrained. This is mostly because of the steepness of the galaxy mass function at such high masses: a small change in the slope results in a large variation in the sparse population of $10^{10} {\rm M}_{\odot}$ black holes. The intrinsic scattering \citep[e.g.][]{grg+09} and poor knowledge of the behavior of the MBH-galaxy relations at the high mass end \citep[e.g.][]{ltrf07} add further uncertainties, making the coalescence rate estimate problematic. As is clear from the figure, we are unable to place any constraints on the physical models mentioned above; however, as our GW sensitivity improves with time, we will begin to place meaningful constraints on physical models pertaining to the coalescence rate of SMBHBs.

\section{Conclusions}
\label{sec:conclusions}

In this paper we have performed various searches for continuous GWs from non-spinning SMBHBs in circular orbits using both frequentist and Bayesian techniques. Specifically, we have run a fixed-noise frequentist and Bayesian pipeline, as well as a varying noise Bayesian pipeline. In the absence of any detections we have placed upper limits on the strain amplitude of continuous GWs as a function of GW frequency. We have also computed a lower limit on the distance to such SMBHBs as a function of sky location, as well as placing constraints on the differential coalescence rate of such SMBHBs. Our sky-averaged upper limits on strain amplitude as a function of frequency are a factor of $\sim 3$ times more constraining than the previously published upper limits \citep{yhj+10} and we see good agreement between all three data analysis methods. Although improving, our limits still lie well above the amplitudes of individual sources produced from several realizations of an optimistic SMBHB population. We have shown that with good estimates of the intrinsic noise we can rule out any sources with luminosity distance $<2$ Gpc and a chirp mass of $\sim 10^{10}{\rm M}_{\odot}$. Unfortunately we are not yet able to place any constraints on predictions for the coalescence rate of SMBHBs obtained from both theory and observations.

Throughout the paper we have made several statements about what is needed for completely \emph{robust} data analysis techniques and what will be required from future PTAs in order to secure a \emph{confident} detection of a continuous GW. These statements can be summarized as follows:
\begin{enumerate}

\item Currently we have no way to confidently separate intrinsic noise in the residuals from any GW that may be present. Therefore, it is necessary to include both noise and GW parameters in any data analysis pipeline that aims to be truly robust. This is not to say that fixed-noise methods should not be used; instead we advocate a hierarchical approach where the faster fixed-noise methods are used as a first-pass and then followed up with a full GW plus noise search. Lastly, a signal with more information, such as that from an eccentric system, could help break this degeneracy between signal and noise models and will be the subject of a future paper. 

\item Even with simultaneous noise and GW characterization, unless we have several well timed pulsars (with very similar timing precision on all) with decent sky coverage, a \emph{confident} detection of a continuous GW is unlikely even if the signal is loud. 

\end{enumerate} 
While not as likely as a detection of a stochastic GW background, with continually improving timing precision, the addition of new pulsars to PTAs and improved data analysis techniques, prospects are good for obtaining astrophysically constraining GW limits, or possibly even a detection of a continuous GW, over the next decade. 

\acknowledgements
The authors would like to thank Neil Cornish and Stephen Taylor for many useful discussions regarding the data analysis methods presented in this work. The NANOGrav project receives support from the National Science Foundation (NSF) PIRE program award number 0968296. All computational work was performed on the Nemo cluster at UWM supported by NSF grant number 0923409. NANOGrav research at UBC is supported by an NSERC Discovery Grant and Discovery Accelerator Supplement. XS and JE were partially funded through an NSF CAREER award number 0955929. JE was partially funded through the Wisconsin Space Grant Consortium.  MV was supported by the Jet Propulsion Laboratory RTD program. Parts of this work were performed at JPL under contract with the National Aeronautics and Space Administration. Data for this project were collected using the facilities of the National Radio Astronomy Observatory and the Arecibo Observatory.  The National Radio Astronomy Observatory is a facility of the NSF operated under cooperative agreement by Associated Universities, Inc. The Arecibo Observatory is operated by SRI International under a cooperative agreement with the NSF (AST-1100968), and in alliance with Ana G. M\'{e}ndez-Universidad Metropolitana, and the Universities Space Research Association. Work at Cornell University was supported by NSF PIRE award 0968126. R.vH. is supported by NASA Einstein Fellowship grant PF3-140116.

\appendix

\section{Gravitational Wave Frequency Evolution}
\label{sec:freq_ev}
Since GWs will radiate power away from a SMBHB source, to compensate for this loss of energy the orbital separation must decrease with time. Equivalently, though Kepler's third law; GW radiation will cause the orbital frequency to \emph{increase} with time. By setting the power radiated in GWs equal to the change of orbital energy due to increasing orbital frequency, $-d E_{\rm orbit}/dt$, we obtain
\be
\dot{\omega} = \frac{96}{5}\mathcal{M}^{5/3}\omega^{11/3},
\ee
where $\mathcal{M} = (m_1 m_2)^{3/5}/(m_1+m_2)^{1/5}$ is the chirp mass and $\omega$ is the orbital frequency of the binary system (note $\omega_{\rm gw} = 2\omega$). We can now use this expression to analytically solve for the orbital frequency as a function of time
\be
\label{eq:worb}
\begin{split}
\int_{t_0}^{t} dt &= \frac{5}{96} \mathcal{M}^{-5/3} \int_{\omega(t=t_0)}^{\omega(t)}\, \, d\omega  \,\omega^{-11/3}\\
t-t_0 &= \frac{5}{256} \mathcal{M}^{-5/3}\lp \omega_0^{-8/3} - \omega(t)^{-8/3} \rp \\
\therefore & \,\, \omega(t) = \omega_0 \lp 1 - \frac{256}{5}\mathcal{M}^{5/3}\omega_0^{8/3}(t-t_0) \rp^{-3/8},
\end{split}
\ee
where $t_0$ is the time at which the first measurement was made on earth and $\omega_0 = \omega(t=t_0)$ is the initial orbital frequency. For a circular orbit, we define the phase to be
\be
\frac{d \Phi}{dt} = \omega.
\ee
We can solve this equation similarly 
\be
\label{eq:phit}
\begin{split}
\int_{\Phi(t=t_0)}^{\Phi(t)} d\Phi & = \int_{t=t_0}^{t} dt^{\prime} \omega(t^{\prime}) \\ 
\Phi(t) - \Phi_0 &= \int_{\omega(t=t_0)}^{\omega(t)}\, \, d\omega  \,\frac{\omega}{\dot{\omega}} \\
&= \frac{5}{96} \mathcal{M}^{-5/3}\int_{\omega(t=t_0)}^{\omega(t)}\, \, d\omega\, \omega^{-8/3} \\
\therefore & \,\, \Phi(t) = \Phi_0 + \frac{1}{32\mathcal{M}^{5/3}}\lp \omega_0^{-5/3} - \omega(t)^{-5/3} \rp,
\end{split}
\ee
where, again, $\Phi_0 = \Phi(t=t_0)$.

Eqs. \ref{eq:worb} and \ref{eq:phit} are true in general  and can be applied when the frequency evolves appreciably over the total observing time. However, it is very useful to work under the assumption of slowly evolving binaries where $T_{\rm chirp}\gg T$, with $T$  the observing time and 
\be
T_{\rm chirp}=\frac{\omega_0}{\dot{\omega}}=3.2\times 10^5 \,{\rm yr}\lp \frac{\mathcal{M}}{10^8 \,{\rm M}_{\odot}} \rp^{-5/3} \lp \frac{f_0}{1\times 10^{-8} \,{\rm Hz}} \rp^{-8/3}.
\ee
Since typical PTA observations are on the order of 10 -- 20 years and $T/T_{\rm chirp}\sim 10^{-4}$, this is a safe assumption for a broad range of masses and initial orbital frequencies of interest. With this approximation we can write the orbital frequency and phase for the earth term simply as
\begin{align}
\Phi_e(t)&=\Phi_0+\omega_0(t-t_0)\\
\omega_e(t)&=\omega_0.
\end{align}
However, for the pulsar term we are "seeing'' the phase and frequency at a retarded time $t_p = t - L(1-\cos\mu)$, where $L$ is the pulsar distance and $\mu$ is the angle between the GW and the pulsar on the sky. Because pulsar distances are on the order of a few kpc, this means that the total time baseline is on the order of thousands of years and we would expect frequency evolution over those timescales. However, just because the pulsar "sees'' a different frequency than the earth, this does not mean that the frequency at the pulsar changes over the observation time. For this reason we can write the phase and frequency at the pulsar in a similar manner
\begin{align}
\Phi_p(t)&=\Phi_{p,0}+\omega_pt\\
\omega_p(t)&=\omega_p.
\end{align}
We can determine the "pulsar frequency'' by evaluating Eq. \ref{eq:worb} and setting $t = t_p$
\be
\begin{split}
\omega_p(t) &= \omega_0 \lp 1 - \frac{256}{5}\mathcal{M}^{5/3}\omega_0^{8/3}(t_p-t_0) \rp^{-3/8}\\
&= \omega_0 \lp 1 + \frac{8}{3}\frac{\dot{\omega}_0}{\omega_0}L(1-\cos\mu) + \frac{8}{3}\frac{\dot{\omega}_0}{\omega_0}(t-t_0)\rp^{-3/8}\\
&\approx \omega_0 \lp 1 + \frac{8}{3}\frac{\dot{\omega}_0}{\omega_0}L(1-\cos\mu) \rp^{-3/8} \equiv \omega_p,
\end{split}
\ee
In the above, we can safely ignore the last term in the second line by the reasoning that the frequency does not evolve over the observation time. Notice that the pulsar term frequency is always less than the earth term frequency as we are observing the dynamics of the SMBMB in the past when the orbital separation was larger. Determining the pulsar phase in this approximation is a bit trickier. Re-writing Eq. \ref{eq:phit}, we get
\be
\begin{split}
\int_{\Phi(t_{p,0})}^{\Phi(t_p)} d\Phi &= \int_{t_{p,0}}^{t_p} dt^{\prime} \omega(t^{\prime}) \\
\Phi_p(t) - \Phi_{p,0} & =\omega_p(t-L(1-\cos\mu)) + \omega_pL(1-\cos\mu)\\
\therefore  \Phi_p(t) &= \Phi_{p,0} + \omega_p t,
\end{split}
\ee
where we have used the fact that $\omega(t) = \omega_p$ in the region of integration and we have adopted a notation in which $\Phi_p(t) \equiv \Phi(t_p)$. To determine the initial phase at $t=-L(1-\cos\mu)$ we use Eq. \ref{eq:phit} to obtain
\be
\label{eq:phip}
\Phi_{p,0} = \Phi(t=-L(1-\cos\mu)) = \Phi_0 + \frac{1}{32\mathcal{M}^{5/3}}\lp \omega_0^{-5/3} - \omega_p^{-5/3} \rp
\ee
Although the above expressions for $\Phi_e(t)$ and $\Phi_p(t)$ are approximations, they hold true for nearly all values of $\mathcal{M}$ and $\omega_0$ that we would expect in nature.

\section{Alternative $\mathcal{F}_p$-statistic Derivation}
\label{sec:alternatefstat}

The $\mathcal{F}_p$ statistic was introduced in \cite{esc12} as a continuous GW detection statistic. However, when applied to single pulsars (as opposed to the full PTA) it essentially acts as a noise weighted periodogram. While a specific notation was introduced in the original work we will use a consistent notation here to avoid confusion. In the same manner as above, let the pulsar timing residuals be denoted as
\be
\delta t = M\delta\boldsymbol{\xi} + n + F\mathbf{a},
\ee
where now $n$ encapsulates all noise in the data (parameterized by some parameters $\vec\phi$) and $F\mathbf{a}$ is the Fourier decomposition of a single GW source and \emph{not} the non-white noise components. Here we will assume a flat prior for the Fourier coefficients  $\mathbf{a}$. We now write the log of the marginalized likelihood ratio 
\be
\ln\,\Lambda = \ln\, p(\delta t | \mathbf{a}, \vec\phi) - \ln\, p(\delta t| \vec\phi) = \delta t^T \tilde{C}^{-1} F\mathbf{a} - \frac{1}{2}\mathbf{a}^TF^T\tilde{C}^{-1}F\mathbf{a}
\ee
where, in a similar manner as above $\tilde{C}^{-1}=G\lp  G^T C G \rp^{-1} G^T$ and $C$ is the covariance matrix of the noise in the residuals. Finding the maximum likelihood values of $\mathbf{a}$ and plugging it into the likelihood ratio we arrive at the $\mathcal{F}_p$-statistic for a single pulsar
\be
\mathcal{F}_p = \frac{1}{2}\delta t^T \tilde{C}^{-1} F(F^T\tilde{C}^{-1}F)^{-1}F^T\tilde{C}^{-1}\delta t.
\ee
Note, that we are projecting the noise weighted residuals onto a Fourier basis and then taking the square. Essentially this is a noise weighted power spectrum and is identical to previous expressions for the $\mathcal{F}_p$-statistic. 

\section{MCMC Implementation Details}
\label{sec:mcmc}

Here we will go over the specific implementation of the PTMCMC algorithm. The algorithm is based on that presented in E13 but has been slightly modified to be more efficient and robust. Specifically we will detail the jump proposals used in this work and the setup of the parallel tempering chains and thermodynamic integration calculation.

\subsection{Jump Proposals}
\label{sec:jumps}

In order to facilitate good mixing of the MCMC chains, especially in large parameter spaces, it is very important to have good jump proposals. In our implementation of the PTMCMC algorithm we use a jump proposal that is composed of a randomized cycle of sub-proposals. Here we will briefly outline the different jump proposals used in the cycle. 

\subsubsection{Correlated Jumps}

As described in E13, we use an Adaptive Metropolis (AM) scheme to make correlated jump proposals. In essence, this jump uses the previous points in the chain's history to construct a sample covariance matrix, $C_n$ \citep{hst01}, that is then updated throughout the run. In practice we update this covariance matrix every 1000 iterations. Primarily the sample covariance matrix is multiplied by a scale parameter $s_d=2.4^2/n_{\rm dim}$, which gives an optimal 25\% acceptance rate in the case of Gaussian posterior distributions; however, we will occasionally make small jumps (scale by 0.01) or large jumps (scale by 10). This jump is used in $\sim$ 20\% of our total jump cycle.

In large parameter spaces, as we encounter when modeling the GW and noise simultaneously, the above method can result in very low acceptance and thus, slow convergence. \cite{hst05} introduce the Single Componentwise Adaptive Metropolis (SCAM) algorithm in which only one \emph{uncorrelated} variable is updated in the jump proposal. If the variables are completely uncorrelated, then this method is identical to using the AM algorithm but only updating one parameter. However, if the parameters are correlated, as they are in our problem, we can define a set of uncorrelated parameters
\be
y = U^Tx,
\ee
where $x$ is our original vector of parameters and $U$ is defined by the eigenvalue decomposition $C_n=USU^T$, where $U$ is a unitary matrix and $S$ is diagonal. It is then easy to see that the covariance matrix of $y$, averaged over many steps in the chain is
\be
\langle y y^T\rangle = U^T\langle x x^T\rangle U = U^T U S U^TU = S.
\ee
Since $S={\rm diag}\{\sigma_s^2\}$ is a diagonal matrix, each $y$ represents an uncorrelated parameter.  Therefore, we choose an uncorrelated parameter at random and propose the jump
\be
y_{i+1}^j = y_i^j + 2.4\mathcal{N}(0,\sigma_s^j),
\ee
where $N(0,\sigma_s)$ is a zero mean Gaussian deviate with variance $\sigma_s^2$, $i$ is the iteration number, and $j$ is the parameter number. We can then relate the jump in the uncorrelated parameters back to a jump in the correlated parameters
\be
x_{i+1} = Uy_{i+1}.
\ee
If U is not the identity matrix (i.e., the parameters, $x$, are correlated) then this means that we will jump in combinations of correlated physical parameters even though we only jump in \emph{one} uncorrelated component at a time. We have found that jumps of this kind greatly improve mixing when running with a large number of search parameters (e.g. $>$100). This jump is used in $\sim$ 40\% of our total jump cycle.

We also employ a third type of correlated jump proposal known as differential evolution (DE) \citep{b06}. Differential evolution is a simple genetic algorithm that also makes use of the previous history of samples in the chain. A differential evolution jump can be constructed as follows. First choose, at random, two previous iterations of the chain. Denote the parameter vector at those two new points as $x_m$ and $x_n$. The DE jump is then
\be
x_{i+1} = x_i + s_{DE}(x_m-x_n),
\ee
where $s_{DE}$ is a scale factor which we choose to be $s_{DE}=2.4^2/n_{\rm dim}$ and $s_{DE}=1$, each with 50\% probability. The first scale factor here is identical to that used in the AM jumps and the second is known as a ``mode jump'', that is, if $x_m$ and $x_n$ are located at two different modes of the posterior distribution, then the mode jump will result in a jump that stays on the same mode as $x_i$ or jumps to the other mode. For this reason, DE jumps are usually employed if there are strong mulimodal structures in the posterior pdf, which we may expect in the case of a weak continuous GW. Also, since we are drawing points from the posterior, then these jumps will also ``learn'' about any correlations among parameters and will be taken into account in the jump proposal. This jump is used in $\sim$ 20\% of our total jump cycle.

Finally, we use a special jump proposal for the chirp mass and luminosity distance. This jump makes use of the fact that there is likely to be a large correlation between chirp mass and luminosity distance for nearly non-evolving sources. If the frequency evolution is negligible then the waveform only has information about the combination $\mathcal{M}^{5/3}/d_L$, with a very weak mass dependence in the frequency derivative term. For this jump, we draw a random luminosity distance value from the prior and then resolve for what value of chirp mass will keep the combination $\mathcal{M}^{5/3}/d_L$ constant. This jump is used in $\sim$ 5\% of our total jump cycle.

\subsubsection{Uncorrelated Jumps}

Although we use mostly correlated jump proposals, about 15\% of our jumps consist of uncorrelated jumps. In many cases, these uncorrelated jump proposals are simple draws from the prior distribution. For prior draws, we have four different jump proposals. Since all pulsars will have a strong white noise component but a weak red noise component and likely no visible GW signal, we draw from the white noise, red noise and GW prior distributions separately with different weights. Red noise and GW (including the pulsar distance) prior draws account for $\sim$ 12\% of our total jump cycle. Finally, we also occasionally make white noise and full parameter space prior draws, which account for $\sim$ 3\% of our total jump cycle. Although this is quite a large percentage of jumps that draw from the prior it greatly improves mixing in our case when we have many search parameters with broad posterior distributions. It should also be noted that when doing injections, we reduce the amount of prior draws to about $\sim$ 5\% of the total jump cycle.

\subsubsection{Auxiliary Pulsar Mode Jump}

In E13, we discussed the difficulty posed by including the pulsar distance as a search parameter, showing that a very small change to the pulsar distance ($\le$ 1 pc) can result in a phase shift in the GW waveform of order $2\pi$. In that work we sidestepped this problem by breaking the pulsar term into a ``phase'' term and an ``evolution'' term. The phase term corresponds to very small jumps in the pulsar distance that will change the constant phase of the pulsar term, whereas the evolution term corresponds to large jumps in the pulsar distance that will change the frequency evolution. We used separate parameters to jump in the phase and evolution. More explicitly, we introduce a pulsar phase for each pulsar that is used in the phase term and \emph{also} include the pulsar distance that is only used in the evolution term. While this method allows for good mixing and acceptance rates, it adds an extra $N_{\rm psr}$  parameters to the search.

Here we will describe a new method that does not require any additional parameters. This jump technique is summarized as follows:
\begin{enumerate}

\item Perform initial jump (either correlated or uncorrelated as described above).

\item Construct pulsar phase of Eq. \eqref{eq:phip} using the new parameters. This phase is likely several radians from the pre-jump pulsar phase due to the pulsar distance jump.

\item We desire a small Gaussian jump in the pulsar initial phase. To accomplish this we will slightly modify the pulsar distance such that
\be
\Phi_{p,0}(L^1+\delta L) = \Phi_{p,0}^0 + \delta \phi,
\ee
where the $1$ and $0$ superscripts denote post and pre-jump values, respectively, $\delta L$ is a small pulsar distance offset, and $\delta \phi$ is a small Gaussian phase jump. We can re-write the above expression
\be
\Phi_{p,0}^1 + \frac{d\Phi_{p,0}}{dL}\bigg|_{L=L^1}\delta L = \Phi_{p,0}^0 + \delta \phi,
\ee
where $\Phi_{p,0}^1=\Phi_{p,0}(L^1)$ and we have simply used a Taylor expansion. Making use of Eq. \eqref{eq:phip} we solve for $\delta L$
\be
\delta L = \frac{\Phi_{p,0}^1 - \Phi_{p,0}^0 + \delta \phi}{\omega_p(1-\cos\mu^1)}.
\ee

\item Now let $L_{\rm new} = L^1 + \delta L$.

\end{enumerate}

Essentially what we have done is to turn a pulsar distance jump into a pulsar phase jump. So in essence we are not breaking detailed balance as we are simply using the pulsar distance as an \emph{auxiliary} parameter and initial pulsar phase as the actual \emph{search} parameter. This auxiliary jump is called after every jump proposal in the cycle to ensure reasonable acceptance rates.

\subsection{Parallel Tempering and Evidence Evaluation}
\label{sec:PTev}

Here we will describe our parallel tempering and thermodynamic integration techniques used to calculate the Bayesian evidence.  We want our algorithm to then quickly locate the global maxima in the parameter space. To accomplish this in a way that satisfies detailed balance we make use of parallel tempering. This technique involves different chains exploring the parameter space simultaneously, each with a different target distribution
\be
p(\vec\phi|d,\beta)=p(\vec\phi)p(d|\vec\phi)^{\beta},  
\ee 
where $\beta\le1$ is the inverse "temperature". This will essentially flatten out the likelihood surface allowing the chains to more freely explore the entire prior volume. The "hot" chains will inform the "colder" chains and vice versa by proposing parameter swaps between different temperatures. A parameter swap between the $i$th and $j$th temperature is accepted with probability $\alpha=\min(1,H)$, where the multi-temperature Hastings ratio is
\be
H_{i\rightarrow j}=\frac{p(d|\vec\phi_i,\beta_j)p(d|\vec\phi_j,\beta_i)}{p(d|\vec\phi_i,\beta_i)p(d|\vec\phi_j,\beta_j)}.
\ee
By swapping parameter states between different temperatures this ensures rapid location of the global maxima. In practice, we perform swaps only between adjacent temperature chains every 1000 iterations. The true posterior samples will come from the $\beta=1$ chain but the higher temperature chains can be used to evaluate the evidence via thermodynamic integration (see e.g. \cite{lc09} and references therein). Consider the evidence for a chain with temperature $1/\beta$ as part of a partition function
\be
\begin{split}
Z(\beta)&=\int d\vec\phi\, p(d|\vec\phi,\mathcal{H},\beta)p(\vec\phi|\mathcal{H})\\
&=\int d\vec\phi\, p(d|\vec\phi,\mathcal{H})^{\beta}p(\vec\phi|\mathcal{H}).
\end{split}
\ee
Since the prior is independent of $\beta$, we can take the log and integrate over $\beta$ to obtain
\be
\label{eq:thermoInt}
\ln\,p(d|\mathcal{H})=\int_0^1d\beta\, \langle \ln\, p(d|\vec{\theta},\mathcal{H}) \rangle_{\beta},
\ee
where $\langle \ln\, p(d|\vec{\theta},\mathcal{H}) \rangle_{\beta}$ is the expectation value of the likelihood for the chain with temperature $1/\beta$. The expectation values are calculated over the post burn-in chains. 

In practice, it is important to choose a temperature ladder such that we explore the entire likelihood surface and recover the full integrand of Eq. \ref{eq:thermoInt}. Here we will closely follow \cite{lc10} in the construction of our temperature ladder and diagnostic techniques. In constructing a temperature ladder to be used with thermodynamic integration it is important to understand that there are two regimes that we are interested in (at least in the GW detection problem). The first regime is the range of temperatures in which the (tempered) likelihood is still in ``contact" with the GW, that is, the data still inform on the GW parameters. Since this is where the bulk of the integrand is concentrated when a signal is present it is very important that we choose a fine temperature spacing here to resolve the point at which the likelihood loses contact with the GW. To do this we choose a geometrically spaced temperature ladder with temperature spacing 
\be
\Delta T = 1 + \sqrt{\frac{2}{n_{\rm dim}}},
\ee
where $n_{\rm dim}$ is the number of dimensions in our search. Now that a temperature spacing is defined we must choose a maximum temperature $T_{\rm max}$ for this regime. This choice is based on the expected maximum SNR of a GW signal in the data. Since $\rho \propto \sqrt{\ln\,p(\delta t | \vec\lambda, \vec\phi)}$, the effective SNR for a chain at temperature $T$ is $\rho_{\rm eff}\propto \rho/\sqrt{T}$, therefore, for a chosen maximum SNR we have
\be
T_{\rm max} = \lp\frac{\rho_{\rm max}}{\rho_{\rm eff,max}}\rp^2,
\ee
where $\rho_{\rm eff,max}$ is the SNR at which we lose contact with the GW signal. For this work we have chosen $\rho_{\rm max}=10$ and $\rho_{\rm eff,max}=3$, corresponding to $T_{\rm max}=11.1$. These values were chosen based on the fact that if we indeed have a signal with $\rho \approx 10$, it would have likely been detected in previous PTA datasets, and $\rho_{\rm eff,max}$ is chosen based on trial and error and simulations. 

If we were only interested in parameter estimation then we would cut off the temperature ladder here; however, for evidence evaluation we must explore the full parameter space. This is the second temperature regime of evidence evaluation via thermodynamic integration. This is most important when varying the noise parameters as well as the GW parameters. Here, we must choose an overall maximum temperature such that we are effectively sampling from the prior. In other words, the temperature must be sufficiently high such that the average log-likelihood has become constant with respect to increasing temperature. In this regime we choose a more coarse temperature spacing with $\Delta T = 1.5$ and geometric spacing. As noted in \cite{lc10} a good diagnostic to ensure that we are using a high enough temperature is to plot the mean log-likelihood for each temperature chain vs. $\beta$.
\begin{figure*}
  \begin{center}
  \includegraphics[scale=1.0]{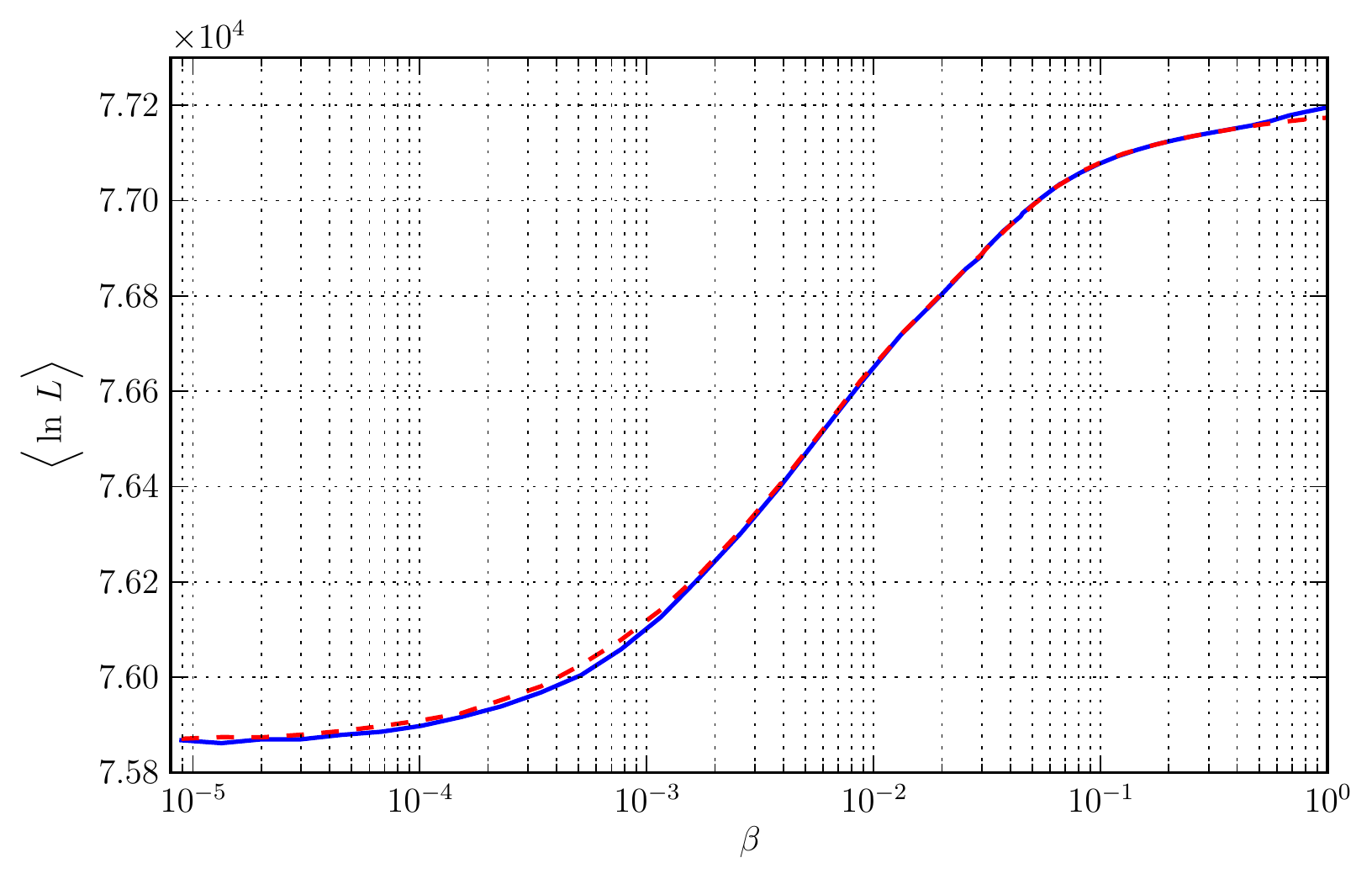}
   \end{center}
  \caption{Mean log-likelihood vs. $\beta$ for GW plus noise ( gray(blue)) and noise (black(green)) models. Here we see that we have indeed explored a sufficient range of temperatures based on the fact that both curves become constant at small $\beta$.}
  \label{fig:logl_vs_beta}
\end{figure*}
In Figure \ref{fig:logl_vs_beta} we plot the mean log-likelihood vs. $\beta$ for GW plus noise (gray(blue)) and noise (black(green)) models. First we notice that at low temperatures (high $\beta$) the GW plus noise model fits the data better based on the higher likelihood values (the data has an SNR 10 GW injection) but that it has slightly lower values at high temperature (low $\beta$) because of the expanded prior volume due to the GW parameter space. Since the Bayesian evidence is the area under these curves the question that is being answered by computing a Bayes factor is ``Does the fact that the GW plus noise model fits the data better (low temperature regime) overcome the fact that that model has a larger prior volume (high temperature regime)?''. Because of this it is crucial that we include temperatures high enough so that the average log likelihood becomes constant, indicating that we are sampling the prior distribution. For this work, we find that a maximum temperature of $\sim 10^5$ is sufficient.

\newpage
\bibliographystyle{apj}
\bibliography{apjjabb,bib}

\end{document}